\documentclass{aastex6}

\usepackage[utf8x]{inputenc}
\usepackage{natbib}
\usepackage{graphicx}
\usepackage{multirow}
\usepackage{mathrsfs,amssymb}
\usepackage{amsmath}

\newcommand       \AU           {\,{\rm AU}}

\newcommand       \ms           {\,{\rm m/s}}

\newcommand       \pc           {\,{\rm pc}}

\newcommand       \Myr          {\,{\rm Myr }}

\newcommand       \Msun         {\,{M_\odot}}

\newcommand{\fljykms}{\mbox{$F_l({\rm Jy-km\ s^{-1}})$}}
\newcommand{\jykms}{\mbox{${\rm Jy-km\ s^{-1}}$}}
\newcommand{\mud}{\mbox{$\mu_{\rm D}$}}
\newcommand{\muh}{\mbox{$\mu_{\rm H}$}}
\newcommand{\mh}{\mbox{$m_{\rm H}$}}
\newcommand{\dpc}{\mbox{$D_{\rm pc}$}}
\newcommand{\bghz}{\mbox{$B_{\rm GHz}$}}
\newcommand{\eup}{\mbox{$E_{\rm up}$}}
\newcommand{\fiso}{\mbox{$f_{\rm iso}$}}
\newcommand{\fc}{\mbox{$f_{\rm C}$}}
\newcommand{\fco}{\mbox{$f_{\rm CO}$}}
\newcommand{\fdust}{\mbox{$f_{\rm dust}$}}
\newcommand{\jj}[2]{\mbox{$J = #1\rightarrow#2$}}
\newcommand{\msun}{\mbox{M$_\odot$}}
\newcommand{\coo}{$^{13}$CO}
\newcommand{\cooo}{C$^{18}$O}

\begin{document}

\title{Disk masses around solar-mass stars are underestimated by CO observations}

\author{Mo Yu\altaffilmark{1}}
\author{Neal J. Evans II\altaffilmark{1, 2}}
\author{ Sarah E. Dodson-Robinson\altaffilmark{3}}
\author{Karen Willacy\altaffilmark{4}}
\author{ Neal J. Turner\altaffilmark{4}}

\altaffiltext{1}{Astronomy Department, University of Texas, 2515 Speedway, Stop C1400, Austin, TX 78712, USA}
\altaffiltext{2}{Korea Astronomy and Space Science Institute, 776, Daedeokdae-ro, Yuseong-gu, Daejeon, 34055, Korea}
\altaffiltext{3}{University of Delaware, Department of Physics and Astronomy, 217 Sharp Lab, Newark, DE 19716  }
\altaffiltext{4}{Mail Stop 169-506, Jet Propulsion Laboratory, California Institute of Technology, 4800 Oak Grove Drive, Pasadena, CA 91109}

\begin{abstract}

Gas in protostellar disks provides the raw material for giant planet formation and controls the dynamics of the planetesimal-building dust grains. Accurate gas mass measurements help map the observed properties of planet-forming disks onto the formation environments of known exoplanets. Rare isotopologues of carbon monoxide (CO) have been used as gas mass tracers for disks in the Lupus star-forming region, with an assumed interstellar CO/H$_2$ abundance ratio. Unfortunately, observations of T-Tauri disks show that CO abundance is not interstellar---a finding reproduced by models that show CO abundance decreasing both with distance from the star and as a function of time. Here we present radiative transfer simulations that assess the accuracy of CO-based disk mass measurements. We find that the combination of CO chemical depletion in the outer disk and optically thick emission from the inner disk leads observers to underestimate gas mass by more than an order of magnitude if they use the standard assumptions of interstellar CO/H$_2$ ratio and optically thin emission. Furthermore, CO abundance changes on million-year timescales, introducing an age/mass degeneracy into observations. To reach factor of a few accuracy for CO-based disk mass measurements, we suggest that observers and modelers adopt the following strategies: (1) select the low-$J$ transitions; (2) observe multiple CO isotopologues and use either intensity ratios or normalized line profiles to diagnose CO chemical depletion; and (3) use spatially resolved observations to measure the CO abundance distribution.

\end{abstract}

\section{Introduction}\label{sec: intro}

Solar System formation models that allow the giant planets to form
within observed protostellar disk lifetimes of a few million years
\citep{haisch01} often require density enhancements up to an order of
magnitude above the minimum-mass solar nebula (MMSN) \citep{pollack96,
hubickyj05, thommes08, lissauer09, Dodson-Robinson_2010_UN, dangelo14}.
Indeed, planet accretion may be a fundamentally inefficient process,
with both collisional fragmentation \citep[e.g.,][]{stewart12} and
planetesimal scattering \citep[e.g.,][]{ida04} contributing to mass loss
during solid embryo growth. Yet disk masses inferred from dust emission
in (sub)millimeter often do not reach the MMSN mass of $0.01 M_{\odot}$
\citep{weidenschilling77, hayashi81}, and are more commonly of order 1-10 Jupiter
masses \citep{andrews07, Williams_Cieza_2011}. Reporting on a survey of T-Tauri stars in Lupus, \citet{Ansdell_2016_ALMA_Lupus} suggested that 80\% of the disks had dust-derived total masses of $< 0.01 M_{\odot}$. However, dust-based disk
mass estimates may be systematically low, because dust continuum
observations lose sensitivity to solids that are much larger than the
observing wavelength \citep{Williams_Cieza_2011}. In addition, the standard assumptions that the dust has a single temperature and that the sub-mm emission is optically thin everywhere in the disk may not be correct. Finally, gas masses
may not be related to the dust masses by the usual interstellar ratio of
100. It is essential to measure the gas mass of disks directly.

One such gas mass measurement came from
\citet{Bergin_HD_2013}, who used {\it Herschel} observations of the HD $(J =
1 \rightarrow 0)$ transition to calculate a mass of $0.06 M_{\odot}$, or
6~MMSN, for the disk surrounding TW~Hydra---a surprisingly large mass
given the star age of $\sim 10$~Myr. The HD lines have since been detected in
two more disks \citep{2016ApJ...831..167M}, and are consistent with gas masses of 1-4.7~MMSN (DM~Tau) and 2.5-20.4~MMSN (GM~Aur). Although the GREAT instrument on the far-IR observatory SOFIA\footnote{https://www.sofia.usra.edu/} covers the frequency of the HD transition, it is not sensitive enough to observe HD in nearby disks. In the absence of the capability to observe the HD \jj10\ transition in disks, CO has been the standard tracer of
the gas mass because it is believed to have simple chemistry and to stay
in the gas phase wherever $T > 20$~K in disks around
Sunlike stars \citep{Oberg_2011_CtoO, Qi13_TWHya}, a region that includes the entire vertical column in the inner 30~AU and the warm surface layers of the outer disk. While the emission from
$^{12}$C$^{16}$O is typically optically thick, it has been suggested \citep{VanZadelhoff_2001, Dartois_2003} that rare isotopologues of CO could be used to probe the disk midplanes. \citet{Yu_2016_COchem}
(hereafter Paper 1) have shown that the vertical optical depth of low-J
rotational emission lines of C$^{17}$O is around unity in the inner
$\sim 20$~AU of a 1.5-MMSN disk, meaning observers could see emission from the disk
midplane---where most of the mass is concentrated---using C$^{17}$O lines. Unfortunately, Paper 1 also
revealed some complexities in the CO chemistry that would interfere with
disk mass measurements. First, the CO abundance varies with distance from 
the star within the planet-forming region. Second, the CO/H$_2$ ratio drops to an order of magnitude below the interstellar value well inside the
CO freeze-out radius because of chemical depletion of CO.
Finally, the CO abundance is a function of time,
which introduces an age-mass degeneracy into the interpretation of
the observations.

Recent attempts to calculate gas-to-dust mass ratios using observations of rare CO isotopologues have also revealed problems.
In their survey of disks in the Lupus star-forming region,
 \citet{Ansdell_2016_ALMA_Lupus} found gas-to-dust ratios, calculated
assuming a constant CO/H$_{2}$ ratio of 10$^{-4}$, to be much lower than
the interstellar value of 100. Yet the stars in the Lupus sample are still accreting, indicating that abundant gas is present. Studies of a more massive disk and star
(2.3 \msun) also found very low gas-to-dust ratios
\citep{2016MNRAS.461..385B}.
Correcting for the joint effects of freeze-out and isotope-selective photodissociation still does not bring the estimated gas-to-dust ratios up to 100, according to the chemical models of \citet{Miotello_2017}. Either the disks are in the
process of dispersing---unlikely given the rapid depletion timescale of
$\sim 10^5$~years once photoevaporation dominates the disk dynamics
\citep[e.g.,][]{hollenbach00, alexander14, gorti16}---or additional pathways to remove gaseous CO (hereafter referred to as chemical depletion) become important in disks, as suggested by \citet{dutrey03}, \citet{favre13}, \citet{Miotello_2017}, and \citet{2016ApJ...831..167M}. Some disks around HAeBe stars also appear to have CO/H$_2 < 10^{-4}$ \citep[e.g.][]{chapillon08, bruderer_2012}.

Also challenging the assumption of simple CO chemistry and a constant
CO/H$_{2}$ ratio in regions where CO is not frozen out are the TW~Hya
observations of \citet{Schwarz_TWHya_2016}, which show a drop in CO
column density at $\sim 20$~AU, while the dust column density remains
roughly constant with radius. In Paper 1, we found that CO chemical depletion due
to dissociation by He$^+$ and subsequent complex organic molecule (COM)
formation causes the CO/H$_2$ abundance ratio to
drop far below the interstellar value of $10^{-4}$ at $r > 20 \AU$, well inside the 
CO freeze-out radius in the model (see figure \ref{fig: CO_contour}). 
For disks with similar chemistry to the Paper 1 models, using interstellar abundance ratios to extrapolate from CO to H$_2$ column density would result in large underestimates of disk mass.

In this work we use radiative transfer models of CO rotational emission to assess the usefulness of rare CO
isotopologues as disk mass indicators, given the possibility of chemical
CO chemical depletion.
We summarize the key results from our chemical evolution models in
section \ref{sec: DiskModel}, including a new model for a more massive disk. 
We present the setup for and results of our line
radiative transfer models in section \ref{sec: lineRTmodel}. In section 
\ref{sec:measure_mass}, we demonstrate that standard gas-mass measurement methods---based on integrated fluxes of lines assumed to be optically thin---fail when applied to the simulated CO emission from our model disk. We also evaluate the performance 
of published optical-depth correction methods \citep{Williams_Best_2014, Miotello_2016}, showing that they are inadequate for disks with non-uniform CO abundance. In section \ref{sec:agemass} we further 
highlight the age-mass degeneracy problem caused by CO chemical 
depletion. Next we show that (1) combining observations of multiple isotopologues, 
(2) using information on line profiles, and (3) examining the spatial distribution of CO can diagnose 
CO chemical depletion  
(section \ref{sec:results_agemass}). In section \ref{sec: obs},
we explore whether our predicted CO/H$_2$ abundance ratios can increase observed masses up to MMSN or higher when applied to the Lupus disk sample of \citet{Ansdell_2016_ALMA_Lupus}. Section \ref{sec:conclusion} summarizes our results.

\section{Disk Model}\label{sec: DiskModel}

We adopt the chemical-dynamical model from Paper 1 as the basis for our
line radiative transfer models. Paper 1 presented the chemical evolution
of a $0.015\Msun$ disk around a Solar-type star for $3$~Myr. In this
paper, we add a chemical evolution model for a disk that is
twice as massive, at $0.03\Msun$. 
The mass distributions and accretion temperatures of both model disks were presented by \citet{Landry_2013}, and
we find the stellar contribution to the disk heating from the dust radiative transfer code RADMC\footnote{http://www.ita.uni-heidelberg.de/~dullemond/software/radmc-3d/; developed by C. Dullemond}. Because our goal is to measure disk masses in giant planet-forming regions, we focus our study on the region covered in \citet{Landry_2013} -- the inner $70$ AU of the disk. (As the disk ages, the amount of mass in the inner $70$ AU decreases due to viscous spearing and accretion -- all disk masses reported in this paper are calculated as $\int 2\pi r \Sigma ~dr$, where $\Sigma$ is the column density, out to 70 AU.) The chemical reaction network is run locally at each independent (r, z) grid point for $3$ Myr, under the assumption that the chemical reaction timescale is much shorter than the viscous timescale---an assumption that is true for freezeout, desorption, and grain-surface reactions, but which may fail for gas-phase reactions. Each disk gridpoint starts with gas and ice abundances
resulting from a 1 Myr simulation of the chemical evolution of a parent molecular cloud; as a result,
a substantial fraction of the carbon is tied up in CO$_2$ and other ices
at the start of disk evolution.

The chemical evolution models include C, H, O, N based on the UMIST database
RATE06 \citep{Woodall_UMIST_2007}. \citet{Woods_Willacy_2009} extended
the network to include C isotopes, and we included both C and O isotopes
in Paper I. The chemical models follow the chemistry of $588$ species,
$414$ gas-phase and $174$ ices, for $3\Myr$ from the beginning of the
T-Tauri phase. The reaction network contains gas-phase reactions (including those that lead to C and O fractionation),
grain-surface reactions, freezeout, thermal desorption, and reactions
triggered by UV, X-rays and cosmic rays, such as isotope-selective
photodissociation.  To make our simulations computationally tractable, we include only
molecules with two or fewer carbon atoms, which limits the network to 11316 reactions. Some grain-surface
hydrogenation reactions, such as 
$(\mathrm{C}_2 \mathrm{H}_5 \; {\rm ice}) + H \rightarrow (\mathrm{C}_2 \mathrm{H}_6 \; {\rm ice})$,
are not included because they only lead to sinks; we can save computational time by not computing the associated
reaction rate and letting $\mathrm{C}_2 \mathrm{H}_5$ ice be the sink instead of $\mathrm{C}_2 \mathrm{H}_6$ ice. 

The simplifications we have made to the reaction network do not artificially remove carbon from the gas phase. In Paper 1 we demonstrated that using a range of chemical networks, input chemical abundances and dust
properties does not change the evolution of CO abundance significantly.
We take the temperatures and abundances from the fiducial model in Paper
1 as the foundation for this study and refer readers to Paper 1 for
detailed discussions of disk model assumptions. The most debatable assumption from Paper 1 is that abundances in gas
and icy components are inherited from the molecular cloud without modification
as they enter the disk. Indeed, \citet{drozdovskaya16} suggest that disk midplane composition is largely determined by the conditions during cloud collapse. \citet{visser09} find that CO ice formed in molecular clouds desorbs during cloud collapse, though it re-freezes without significant abundance modification where the disk temperature is $< 18$~K.

\begin{figure*}[ht]
\centering
\begin{tabular}{@{}cc@{}}
\includegraphics[width=0.4\textwidth]{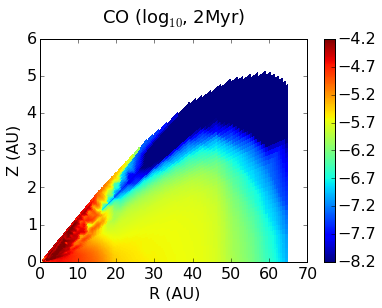} &
\includegraphics[width=0.4\textwidth]{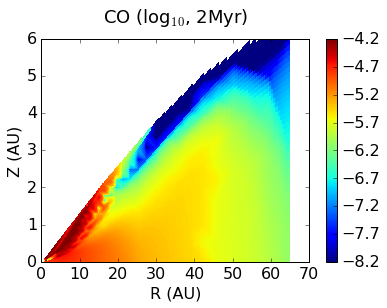} \\
 \end{tabular}
\caption{CO abundance as a function of disk radius (R) and height (Z) at $2$ Myr. We show results from the $0.015 \Msun$ model on the left and the $0.03 \Msun$ model on the right. The abundance is defined as the number density with respect to the number density of hydrogen nuclei (n$_{\rm H}+2\rm n_{\rm H_{2}}$).}
\label{fig: CO_contour}
\end{figure*}

\begin{figure*}[ht]
\centering
\begin{tabular}{@{}cc@{}}
\includegraphics[width=0.4\textwidth]{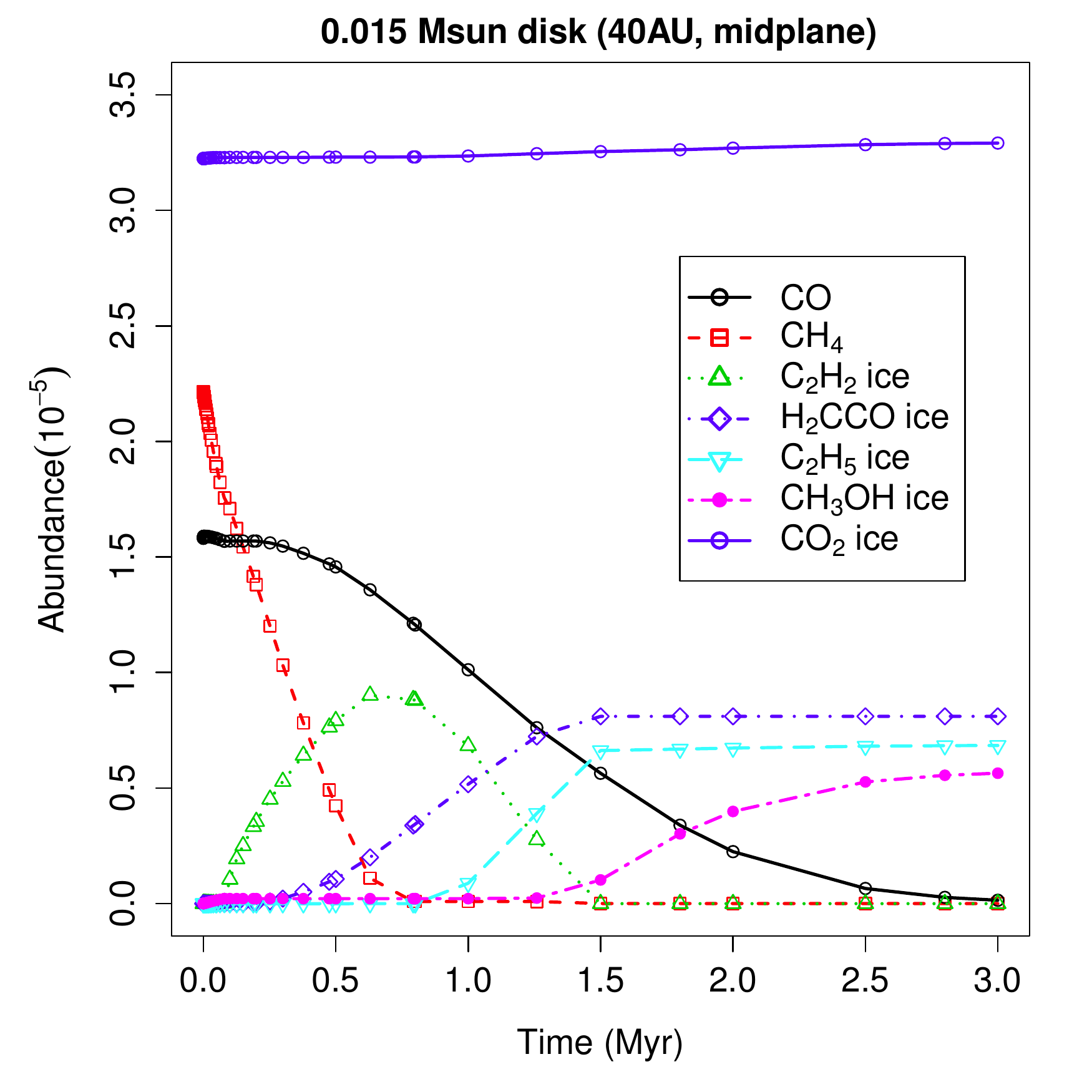} &
\includegraphics[width=0.4\textwidth]{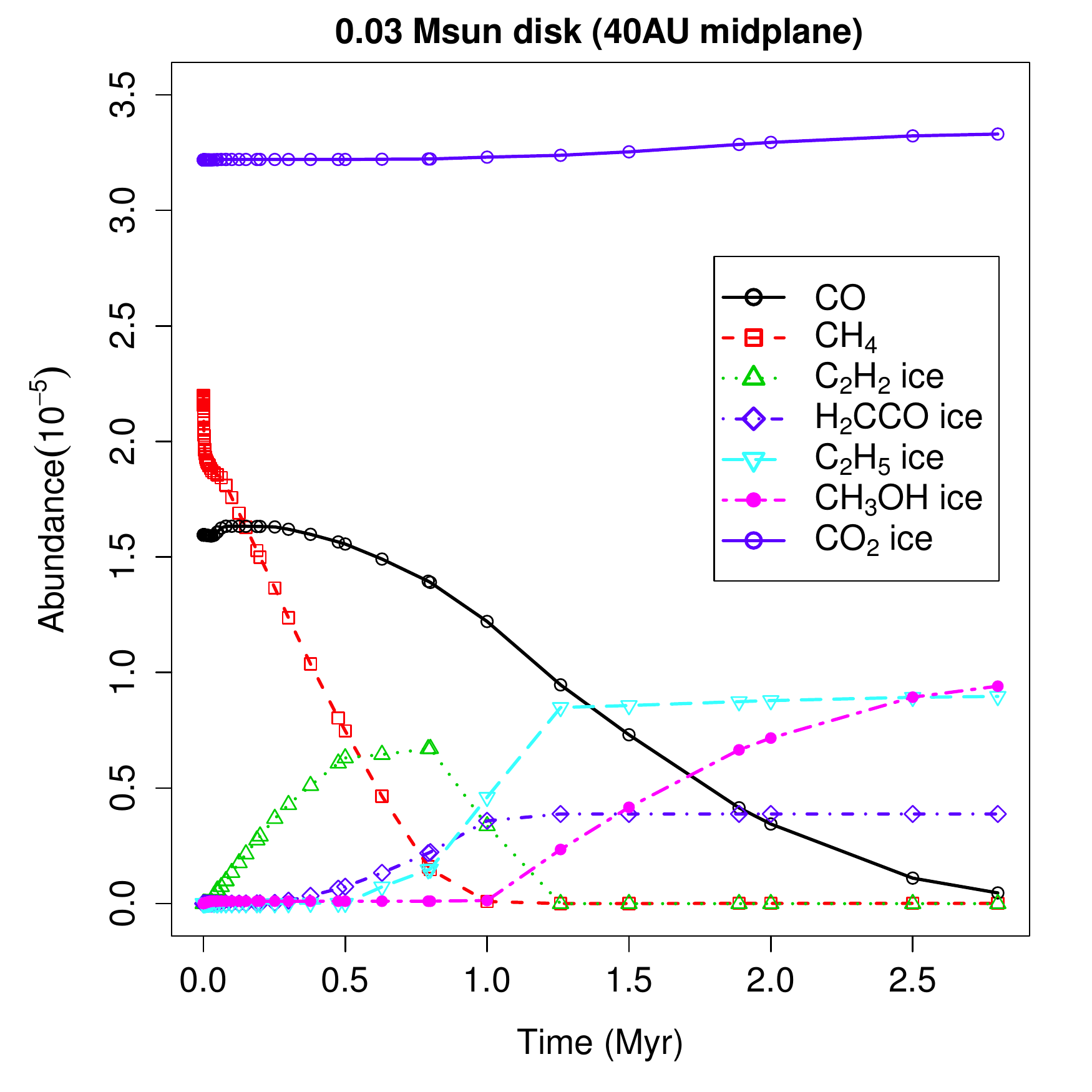} \\
 \end{tabular}
\caption{Abundances of major carbon-bearing molecules as a function of time at $40$ AU on the disk midplane.
We show results from the $0.015 \Msun$ model on the left and the $0.03 \Msun$ model on the right. }
\label{fig: chem_time_evolution}
\end{figure*}

The disk is primarily heated passively from stellar irradiation;
accretion contributes very little to the energy budget
(\citealt{Landry_2013}, Paper 1). 
For the temperature calculation, we use the dust opacities computed by
\cite{Semenov_Henning_2003}, adjusted for our gas/dust ratio. We assume
the dust has already grown and aggregated at the start of the T-Tauri
phase of disk evolution \citep{Oliveira_2010, Perez_2012_AS209,
Birnstiel_graingrowth_2011, Garaud_graingrowth_2013}, and 90\% of the
solids have grown to still larger sizes (pebbles, rocks, etc.), that
contribute very little submillmeter emission. We distinguish these
{\it solids} from {\it dust} and take a
gas/dust mass ratio of 1000. In Paper 1, we demonstrated that the CO
abundance is only weakly sensitive to evolution of the gas/dust ratio
from 100 to 1000. Our grain model assumes that there is a constant
replenishment of small grains by collisions between larger objects (see
\citet{Dullemond_Dominik_2005_smallgrains, Brauer_2008,
Birnstiel_fragmentation_2009, Wada_2008, Wada_2009, Windmark_2012};
though \cite{Zsom_2010} argues against replenishing micron-size dust
grains through collisions) and the size distribution of dust does not
evolve. 
The theory that most of the solid mass has aggregated into pebbles and larger objects
invisible to submillimeter wave observations is the second major, non-standard assumption in our model (though \citet{andrews12} and \citet{cleeves15} modeled the TW~Hya disk with two distinct grain populations, small grains that provide the visible/infrared opacity and large grains with a maximum size of 1~mm that provide the submillimeter opacity). Following \citet{cleeves15}, we assume the 90\% of solid mass that has grown to pebble and larger sizes provides negligible surface area for chemical reactions and exclude it from the chemical model.

As the central star dims during its time on the Hayashi track,
the disk cools and its scale height shrinks. The computational surface of our model
grid, defined as the layer where the optical depth to the disk's own
radiation is $\tau = 0.2$ \citep{Landry_2013}, moves from about two pressure
scale heights to about one scale height above the midplane as the disk
becomes cooler and thinner over time. The chemical model therefore contains fewer grid cells at the end of evolution than at the beginning, as grid layers high above the midplane begin to empty out. In appendix \ref{app: modelpars}, we show that our 
line profile and intensity calculations are only weakly sensitive to the
changing disk surface.

Stellar heating is efficient enough to
prevent CO from freezing out in our modeled region---the inner 70 AU of
the disk---at all vertical heights and at any time in the $3$~Myr of evolution. However, CO is
depleted beyond $20$~AU from the central star due to the formation of
complex organic molecules. 
The CO chemical depletion  is driven by ionization of
helium from X-rays and cosmic rays and happens over a million-year time
scale. As a result, the CO abundance changes both with location in the disk and
with time. We show the color map of CO abundance at $2$ Myr of the disk 
evolution in Fig. \ref{fig: CO_contour} for both the Paper 1 disk
($0.015$ \msun; left) and the new model of the more massive disk ($0.03$ \msun; right).  
The abundance is defined as the ratio of the number density of CO with respect to the 
number density of hydrogen nuclei (n$_{\rm H}+2\rm n_{\rm H_{2}}$). 
After 2~Myr, the CO chemical depletion  front moves inward to 20~AU, 
with the surface layers more depleted than the disk midplane. Fig.\ 
\ref{fig: chem_time_evolution} shows the abundances of major 
carbon-bearing species as a function of time for the midplane at 40~AU in each disk, demonstrating the gradual chemical depletion 
of CO and subsequent sequestration of carbon into organic ices. We note that our network is not extensive enough to determine the 
exact end-product of organic ice formation. The most complex hydrocarbon we include is C$_2$H$_5$, which in reality should hydrogenate
to form ethane, and radicals of the form C$_2$H$_{\rm x}$ should react with carbon atoms or hydrides to form longer carbon chains.
However, these complex products of organic grain surface chemistry are all less volatile than the C$_2$H$_{\rm x}$ species and would
stay on the grain surfaces unless the grains drift radially inward (e.g.\ Birnstiel \& Andrews 2014) or experience a transient
heating event (e.g.\ Cody et al.\ 2017, Cieza et al.\ 2016, Vorobyov \& Basu 2015). Since almost all resolved T-Tauri disks detected
in dust continuum emission have radii much larger than the $\sim 40$~AU C$_2$H$_5$ ice line (Paper 1) in our 
models\footnote{See the catalog of resolved disk images at {\texttt www.circumstellardisks.org}}, we do not expect radial drift
to deposit a significant amount of organic gas that could be recycled to form CO in the inner disk.
To the extent that there are real astrophysical 
disks that evolve quiescently during the T-Tauri phase, our conclusion that carbon liberated by CO chemical depletion becomes sequestered
in ices is robust.
The more massive disk shows a similar pattern of CO chemical depletion to the disk presented in Paper 1, though the depletion timescale is a bit longer because the higher column density decreases the
ionization fraction---and thus the abundance of ionized helium---in the midplane.

The net result of all chemical models presented here and in Paper 1 is that CO becomes severely depleted well inside the CO freeze-out radius in disks with masses above the minimum needed to form planetary systems. Similar effects have been seen in other chemical
evolution calculations (\citealt{Aikawa_1997, 1999ApJ...519..705A, Furuya_carbon_2014, 2014A&A...563A..33W, Bergin14}). \citet{Aikawa_1997} and \citet{1999ApJ...519..705A} first pointed out that CO can react to form less volatile molecules such as CO$_2$, HCN, H$_2$CO, CH$_4$ and larger hydrocarbons over Myr timescales. \citet{Aikawa_1997} used a static disk (not evolving), and \citet{1999ApJ...519..705A} used a vertically isothermal model and found a larger number of simple molecules as products (as opposed to a few complex molecules), but the essence of the chemical depletion of CO is the same as found here and in Paper 1. \citet{2014A&A...563A..33W} simulated the composition of complex organic molecules in a disk with no temperature evolution for about $1$ Myr and found the formation of 
complex organic molecules in the disk midplane via grain-surface reactions, while \citet{Bergin14} found gas-phase organic formation and subsequent freezeout onto grain surfaces.
\citet{Furuya_carbon_2014} investigated the carbon and nitrogen chemistry 
during turbulent mixing, and found that volatile transport enhances COM formation 
near the surface and suppresses it in the disk midplane. It is difficult to compare our results directly with the earlier calculations because our models have different central star properties and less accretion heating, but it is clear that a variety of chemical processes can force a disk's CO/H$_2$ abundance ratio far from the interstellar value.

\begin{figure*}[ht]
\centering
\includegraphics[width=0.4\textwidth]{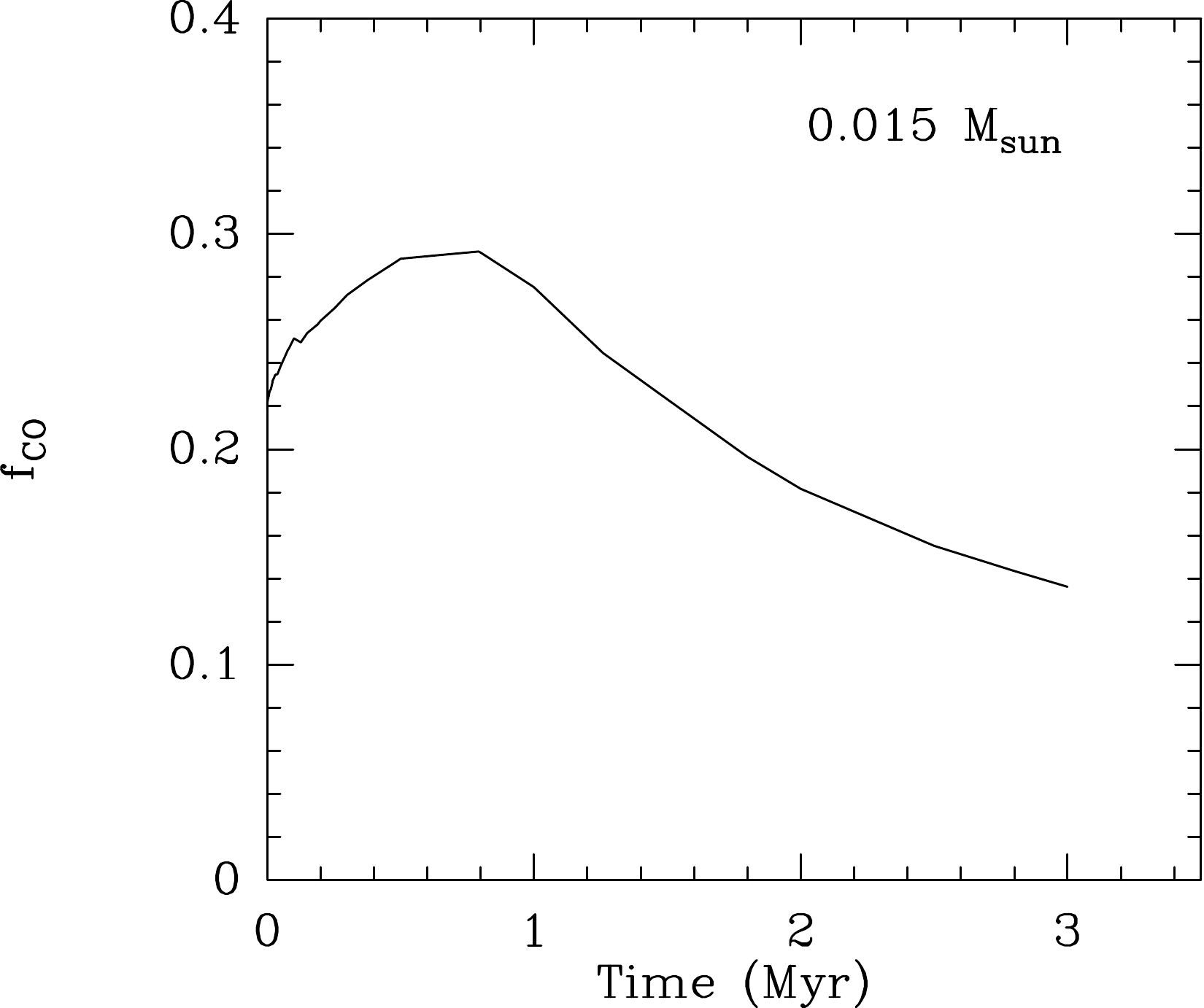}
\includegraphics[width=0.4\textwidth]{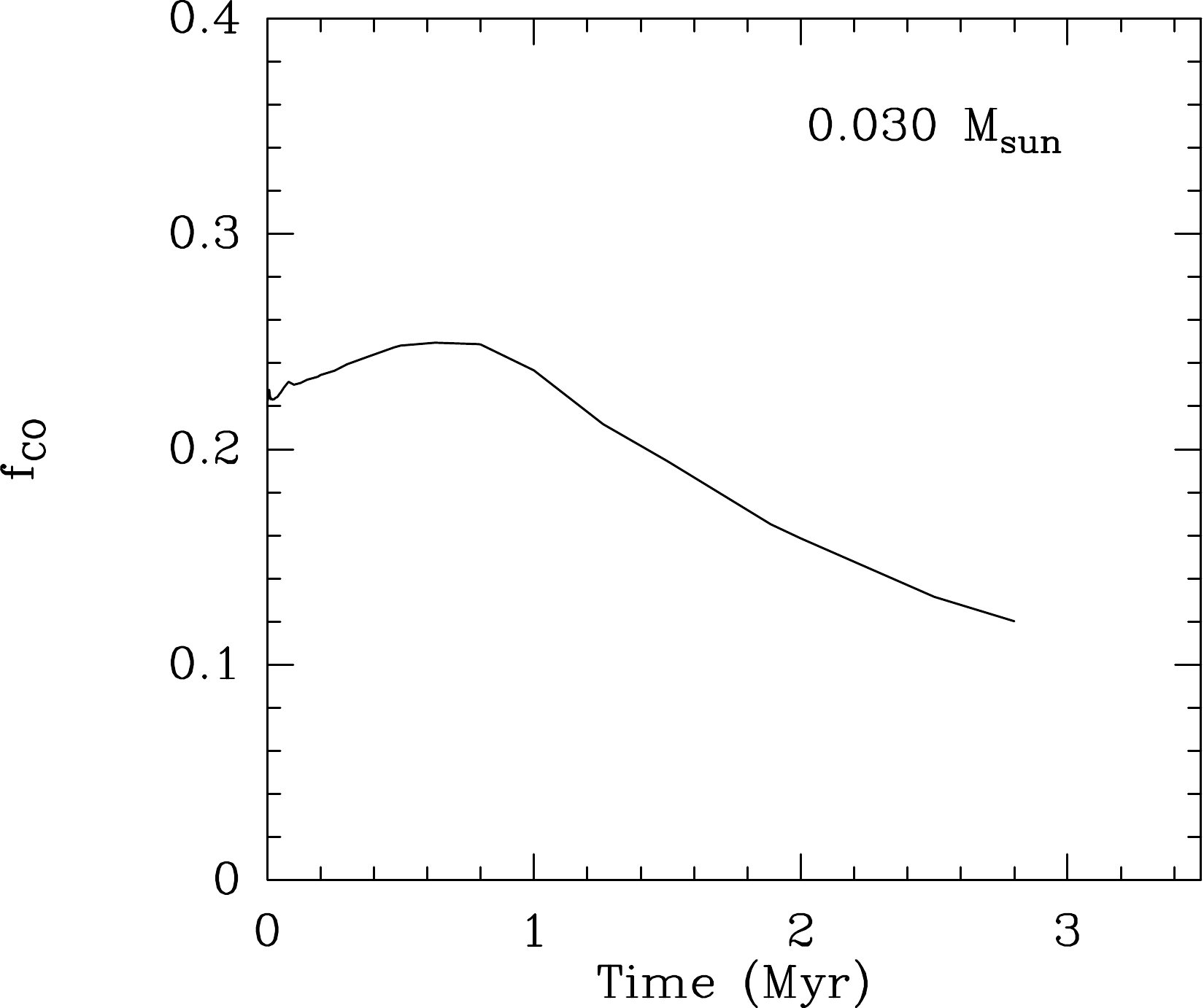}
\caption{
The fraction of C in CO versus time, averaged over the disk, for both the disks, 0.015 \msun\ on the left, 0.030 \msun\ on the right.
}
\label{fcovst}
\end{figure*}

In Figure \ref{fcovst} we show the fraction of our model disk's carbon atoms contained in CO gas (\fco) as a function of time. The fraction is initially low because evolution in the
molecular cloud has sequestered carbon in ices; for a while these evaporate,
increasing \fco, but then CO chemical depletion and formation of icy organics cause
\fco\ to decrease. 
To complicate matters,
the abundance of CO actually increases with time at small radii, as
CO$_2$ ice desorbs and dissociates to form CO gas. Although the $0.015 \Msun$ disk suffers more CO chemical depletion at larger radii, the re-formation of CO gas in the inner disk also proceeds at a higher rate than in the $0.03 \Msun$ disk, leading to a higher disk-averaged \fco\ for the $0.015 \Msun$ disk. These facts will compromise
attempts to measure disk masses using CO isotopes (\S \ref{sec:measure_mass}).
Worse yet, the
CO abundance is a function of time, leading to an age-mass degeneracy in
interpreting observations (\S \ref{sec:agemass}).


\section{Line radiative transfer model and mass estimates}
\label{sec: lineRTmodel}

In order to understand the effects of chemical evolution on CO emission,
we set up radiative transfer models with the publicly available code
LIME \citep[LIne Modeling Engine;][]{Brinch_Hogerheijde_LIME_2010}. LIME
calculates either non-LTE or LTE line excitation and solves the radiative
transfer problem for molecular gas in arbitrary 3D geometries. For
recent examples of studies using LIME, see \citet{Walsh_methanol_2016}
and \citet{Oberg_2015_cyanides}. We adopt the energy levels from the Leiden Atomic and Molecular Database
(LAMDA)\footnote{http://home.strw.leidenuniv.nl/~moldata/}. As a first
order approximation, we do not consider the hyperfine splitting in
C$^{17}$O emission.

As in Paper 1, we model emission from within a radius of $70\AU$ of the central
star, which corresponds to a $1\arcsec$ beam diameter for an assumed
distance of $140\pc$ from the Sun. We consider line broadening due to
Keplerian rotation, thermal velocity and micro-turbulence. Thermal
velocities are calculated assuming a Maxwell-Boltzmann speed
distribution based on the disk's temperature structure from Paper 1. For the
micro-turbulence, we again assume an isotropic Maxwell-Boltzmann speed
distribution with RMS of $100\ms$, consistent with the upper limit to microturbulent speed found by \citet{flaherty15} in a fit to multiple CO emission lines in the HD~163296 disk. For computing the level populations, we
set a minimum scale of $0.07\AU$ to guarantee sub-pixel sampling of both
Keplerian speeds and CO abundance gradients.
We first generate the synthetic datacube of intensity as a function of
$x$, $y$, and velocity for a disk around a $0.95 \Msun$ star at $30
\degr$ inclination, similar to the disk surrounding AS~209.  In velocity
space, the spectra have $300$ channels of $125\ms$ resolution. At any
specific velocity, the synthetic image contains $600 \times 600$ pixels
of $0.003\arcsec \times 0.003\arcsec$ in size. Finally, we generate the
synthetic spectra presented here by integrating each velocity component
over a square with $1.2\arcsec$ sides ($400 \times 400$ pixels), larger
than the angular size of the disk. The pixels not covered by the disk
contribute no flux and are included simply for ease of
integration---here we assume that the sky background contains negligible
flux compared with the disk at all wavelengths.

Our current models assume that the gas temperature is the same as the dust
temperature. This is a valid assumption in estimating the optical depth
of C$^{17}$O and C$^{18}$O, as done in Paper 1, because the emission
primarily comes from disk midplane and the midplane is strongly shielded
from UV radiation. However, hot gas on the disk surface is more
emissive, and we would need to consider the difference between the gas
and dust temperature in order to use our models to fit high-J spectral
lines emitted from the disk surface. Similarly, we would need to revisit
the temperature calculation for a disk surrounding a star with a
stronger UV field, which could decouple the gas and dust temperatures.
In Appendix \ref{app: modelpars}, 
we demonstrate that the LTE approximation is adequate for computing 
the energy level populations, and show that the decreasing disk scale height has little effect on the computed line
profiles (see \S \ref{sec: DiskModel}). 
For our purposes, we have adequately modeled the CO rare isotopologue 
emission, even though our models do not extend vertically to a large number 
of scale heights. 
For the rest of the paper, 
we focus our line profile discussion on \jj32\ 
and \jj21, which are the most commonly observed transitions.

\begin{figure*}[ht]
\centering
\begin{tabular}{@{}ccc@{}}
 \includegraphics[width=0.3\textwidth]{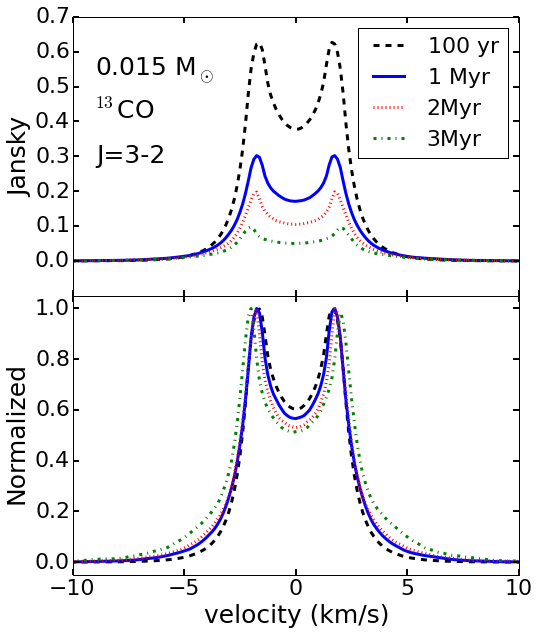} &
  \includegraphics[width=0.3\textwidth]{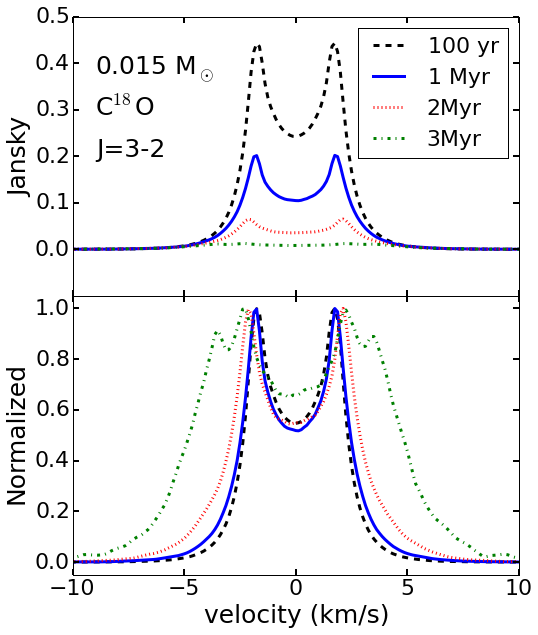} &
  \includegraphics[width=0.3\textwidth]{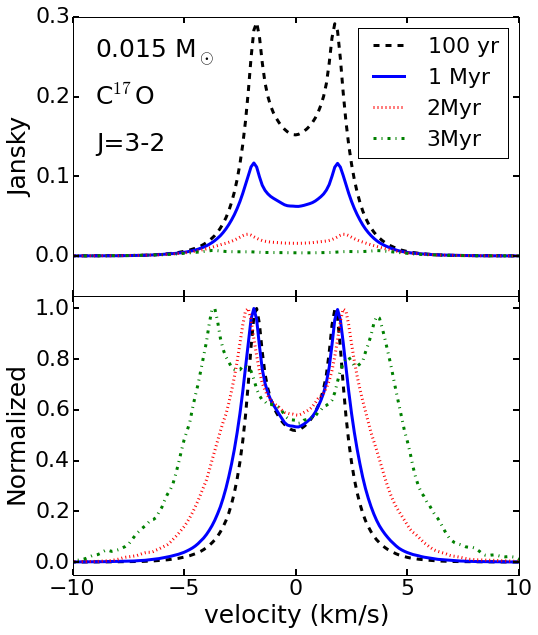} \\
  
  \includegraphics[width=0.3\textwidth]{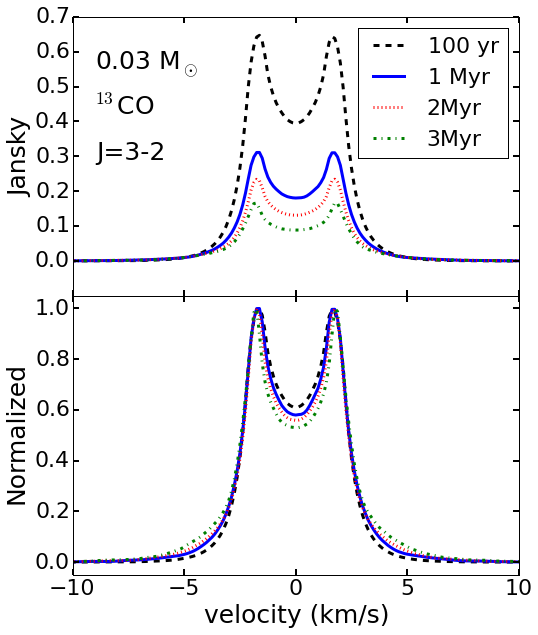} &
  \includegraphics[width=0.3\textwidth]{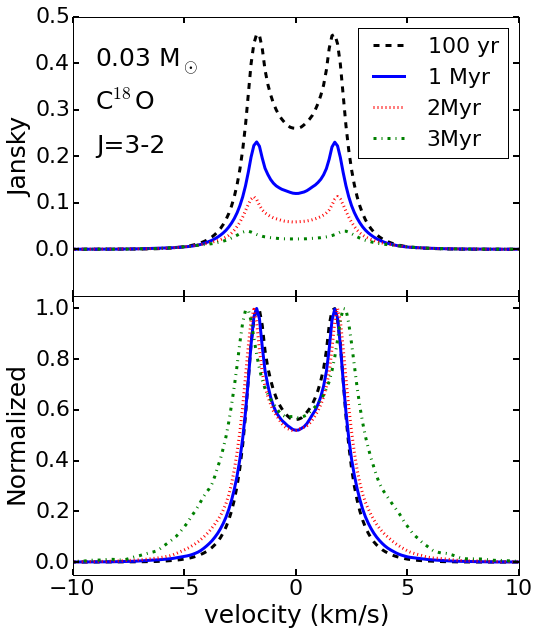} &
  \includegraphics[width=0.3\textwidth]{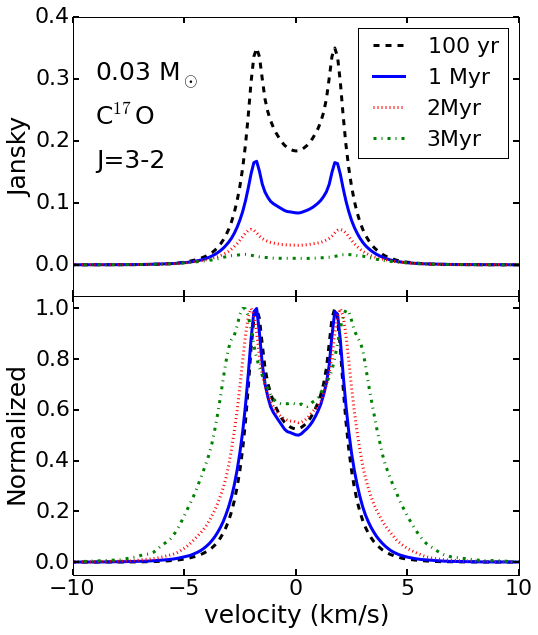} \\
  
 \end{tabular}
  \caption{Time evolution of emission line profiles, with $^{13}$CO on the left, 
  C$^{18}$O in the middle and C$^{17}$O on the right. The top 6 panels show 
  results for the fiducial $0.015 \Msun$ model, and the bottom 6 panels show 
  the results for the comparison $0.03 \Msun$ model. In each group, the top 
  panels show the simulated lines, and the lower panels show the line profiles
  normalized to the peak intensity of each line. The emission becomes
  weaker, and the line profile becomes wider over time for all
  isotopologues, but the change is much more significant for the
  optically thin C$^{18}$O  and C$^{17}$O emission.}
 \label{fig: line_evo}
\end{figure*}

\begin{figure*}[ht]
\centering
\begin{tabular}{@{}cc@{}}
\includegraphics[scale=0.4]{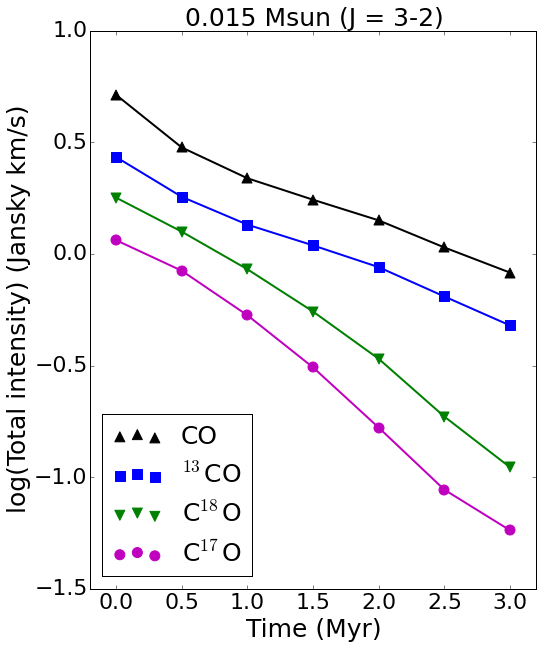} &
\includegraphics[scale=0.4]{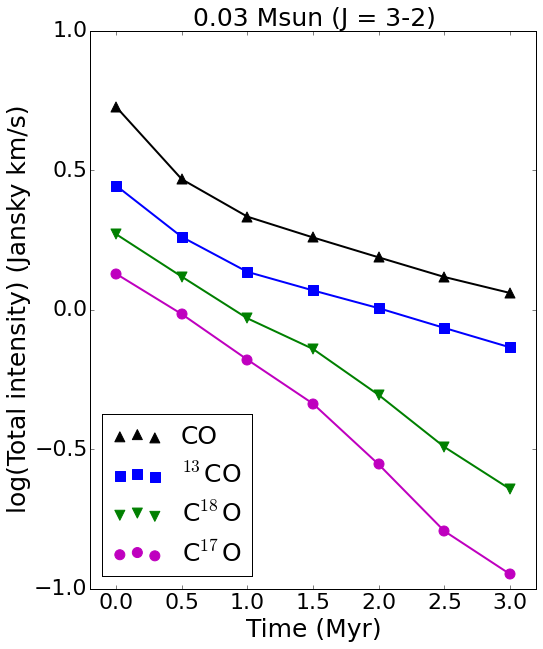}\\
 \end{tabular}
\caption{Total intensities of J=$3-2$ emission from various CO isotopologues 
as a function of time. Results from the fiducial with $0.015 \Msun$ are presented on the left, and those from the fiducial model with $0.03 \Msun$ are presented on the right. In the $0.015 \Msun$ model,  the intensity of $^{13}$CO drops from $2.72 \jykms$ 
by $82\%$ to $0.48 \jykms$  over the $3$ Myr disk evolution, and the intensity 
of C$^{17}$O drops from $1.15\jykms$ by $95\%$ to $0.06 \jykms$.}
\label{fig: time_evolution_intensities}
\end{figure*}


Figure \ref{fig: line_evo} shows one set of results from our line radiative transfer models---the time evolution of the \jj32\ line profiles for three isotopologues and two model disks. Line profiles in Janskys are
plotted as well as {\it normalized} profiles, which better show the evolving
{\it shape} of the lines (see \S \ref{sec: lineprofiles} for more on diagnosing CO chemical depletion  by comparing line profiles from multiple isotopologues). Figure~\ref{fig: time_evolution_intensities} shows the total intensities 
(integrated in velocity space) of \jj32\ emission from all CO isotopologues 
as a function of time. The decline with time is quite dramatic, especially
for the rarer isotopologues where it exceeds an order of magnitude over
3 Myr. The emission from rarer isotopologues is very weak at later times, which can be expensive to observe, especially if one wants to achieve enough signal-to-noise to study the line profiles.

Finally, we use standard observational procedures to try to recover the apparent masses of the disks, as an observer would do. Here we will use
the \jj21\ line, as it is commonly used to obtain disk masses (e.g.,\citealt{Williams_Best_2014}, though
\citet{Ansdell_2016_ALMA_Lupus} use the \jj32\ line). We show in Appendix \ref{sec:rotdiag} that higher $J$ transitions underestimate the
mass more severely than \jj21\ and \jj32. From our model line profiles we
calculate level populations, total number
of CO molecules, and finally gas mass. 
The equations used to derive the number of CO molecules in each
energy level ($\mathcal{N}_J$) from integrated line flux, assuming optically thin emission, are given in Appendix \ref{sec:rotation_diagram_eqs}.
Values of $\mathcal{N}_J$ can then be translated to mass with the following equations.
The total number of CO molecules in the disk $\mathcal{N}$ is given by
\begin{equation}
\mathcal{N} = \frac{\mathcal{N}_J}{g_J} \times Q \times e^{\eup/T}
\label{eq:nco}
\end{equation}
where \eup\ is the upper-state energy  of the
transition in K and $Q$ is the partition function, generally approximated by
$Q = kT/hB$ (with $B$ in Hz). Note that the temperature of the CO reservoir appears in both the exponential and the partition function in Equation \ref{eq:nco}; in Appendix \ref{sec: Testimates} we show that no single temperature describes the CO reservoir.
We then compute the mass of gas from
\begin{equation}
M = \frac{\mathcal{N} \muh \mh \fiso}{\fc \fco}
= 1.668 \times 10^{-53} \msun\ \mathcal{N} \fiso/\fco
\label{masseq}
\end{equation}
where \muh\ is the mean atomic weight of the ISM including He, \mh\ is the mass of a
hydrogen atom, \fiso\ is the ratio of the mass of the most common
isotopologue ($^{12}$C$^{16}$O) to that of the one being used, \fc\ is the abundance ratio of
carbon to hydrogen nuclei, and \fco\ is the fraction of C in CO,
averaged over the disk. $\muh = 1.43$ and our chemical model has $\fc = 7.21 \times 10^{-5}$. 


In the next section we describe how equations \ref{eq:nco} and \ref{masseq} to obtain the correct mass of our model disk.

\section{Even rare CO isotopologues underestimate disk mass}
\label{sec:measure_mass}

$^{12}$C$^{16}$O and $^{13}$C$^{16}$O will be optically thick in disks, so the usual
approach is to observe rarer isotopologues. At first glance, the CO chemical depletion  we predict might be expected to ameliorate optical depth problems, but the concentration of CO to small radii increases optical depths there. At the same time, the
chemical depletion of CO at larger radii causes mass underestimates by lowering the value of \fco.
The net result is that simple analysis fails for the following reasons. 

First, the usual assumption is that $\fco = 1$ inside the CO ice line, but our models challenge
that assumption. Figure \ref{fcovst} shows how \fco\ varies with time,
never rising above 0.3 ($0.015$ \msun\ disk) or $0.25$ ($0.03$ \msun\
disk) and approaching 0.12 by 3 Myr. Since our disk does not include radii
where CO itself would freeze, the low \fco\ is due to what we call
chemical depletion. Alone, with no other error sources in equations \ref{eq:nco} and \ref{masseq}, it will cause underestimates of disk mass
by factors of 3 to 8.

\begin{figure*}[ht]
\centering
\begin{tabular}{@{}cc@{}}
\includegraphics[width=0.4\textwidth]{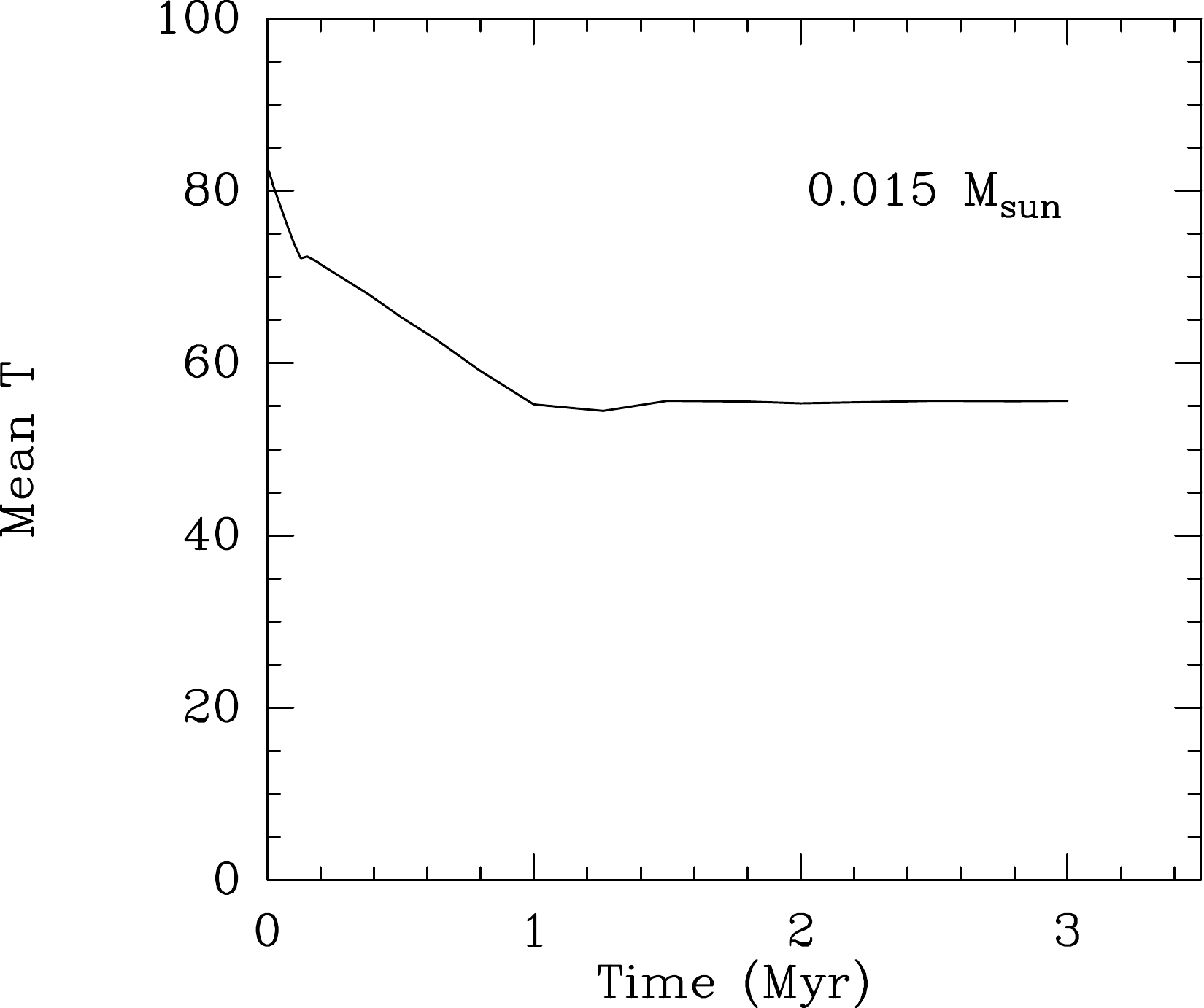} &
\includegraphics[width=0.4\textwidth]{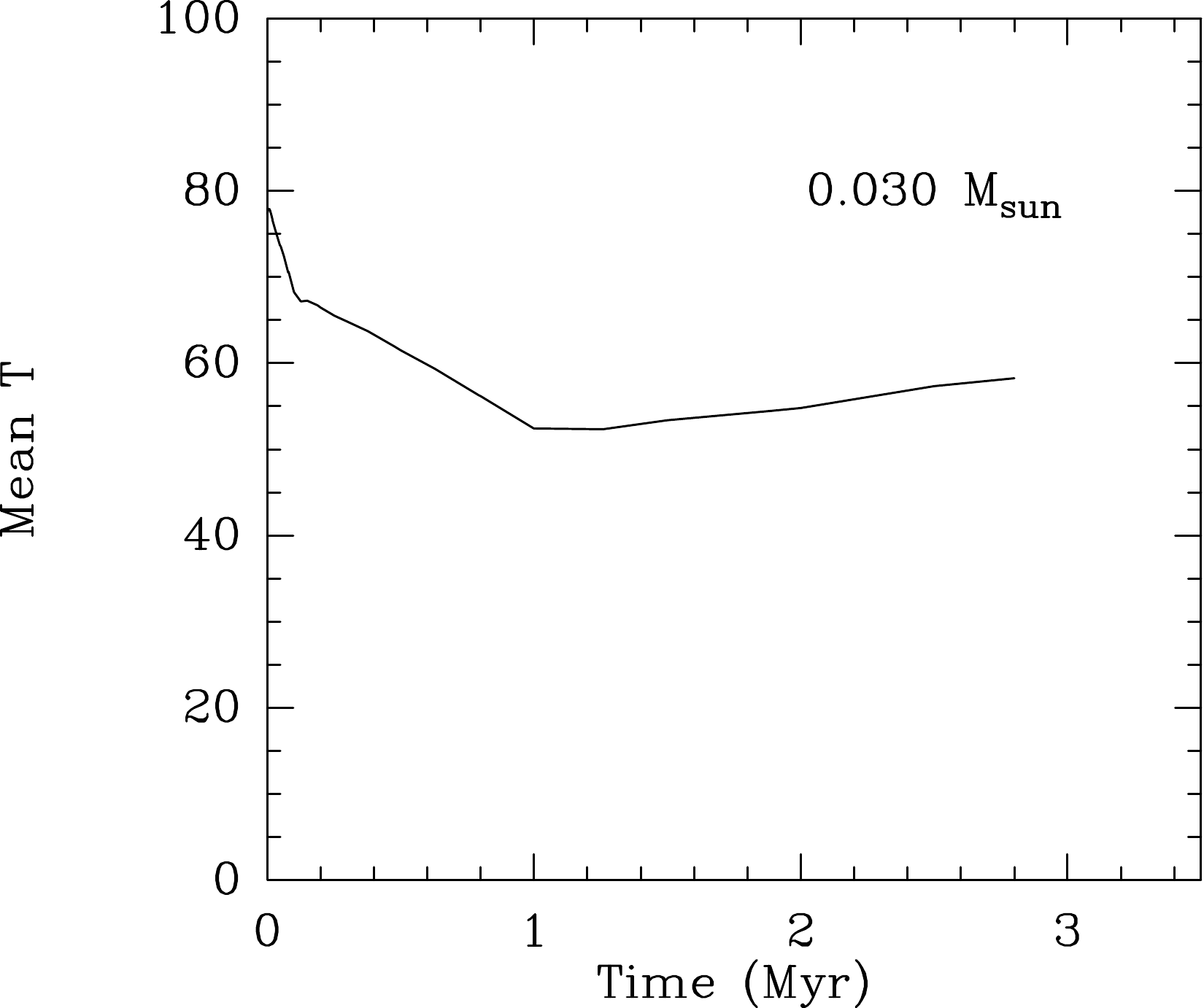} \\
 \end{tabular}
\caption{
The temperature versus time, weighted by the CO density, and averaged over the disk, for both the disks, 0.015 \msun\ on the left, 0.030 \msun\ on the right.
}
\label{fig: Tvst}
\end{figure*}

Second, the concentration of CO toward the inner disk means that
temperatures, and hence partition functions, are higher than usually assumed for much of the CO reservoir. The most
common assumption for temperature is 
$T = 20$ K (e.g., \citealt{Ansdell_2016_ALMA_Lupus}).
A temperature averaged
over the the model disk and weighted by the number density of CO molecules ranges from 50 to 70 K, decreasing as the star and disk
evolve, but stabilizing around 55 K for the 0.015 \msun\ disk after about
1 Myr because the increasing concentration of CO in the inner disk counteracts
the dropping luminosity of the star (Figure \ref{fig: Tvst}).
If CO chemical depletion is operating in the outer disk, assuming $T = 20$ K for the gas where CO is concentrated is never
a good choice; it will systematically underestimate masses by a factor
of about 2. In disks around T-Tauri stars that are less luminous than the proto-sun at 3~Myr, or that receive a low cosmic-ray flux so that cosmic ray-induced photons do not generate CO from CO$_2$ gas in the inner disk, 20~K may be appropriate, but we recommend constructing model-based temperature estimates for the CO reservoir rather than making any assumptions.
As a further complication, we show in Appendix \ref{sec:rotdiag}
that rotation diagrams from multiple transitions are not effective
in determining the best temperature to assume.

Third, the increase in CO abundance in the inner disk causes optical depth
effects where most of the CO actually resides (see Fig. 10 of Paper 1). 
Even if we correct for
the mean \fco\ and somehow find an accurate value of $T$ averaged over the CO reservoir, the standard analysis
still underestimates the mass.
Figure \ref{massplot}
 shows the mass estimates from different isotopes and analysis
procedures versus time, along with the actual disk mass, which decreases
slightly
due to disk spreading and accretion onto the star. The estimates with
black symbols are computed with equations \ref{eq:nco} and \ref{masseq} and based on the assumption of optically thin emission
in the \jj21\ line of the listed isotopologues. They all use the correct \fco\ and the mass-weighted CO temperature for our model---information that would not be available for astronomical sources---yet
they all still underestimate the gas mass substantially. The rarer isotopologues
perform best, suggesting that optical depth is the main culprit.
The recent analysis of the Lupus disks \citep{Ansdell_2016_ALMA_Lupus} by
\citet{Miotello_2017} also suggests optical depth effects, even
after ratios of \coo\ and \cooo\ were used to correct for optical depth. 
The apparent gas to dust
ratio declines with increasing dust mass in their Figure 6; if the dust
mass is a reasonable proxy for disk mass, a more massive disk will have
larger (and more complicated) effects from optical depth. Our results
indicate that models that do not account for the radial dependence of
\fco\ will underestimate the gas mass more severely for more massive disks, as
observed in Figure 6 of \citet{Miotello_2017}.

\subsection{Correcting for optical depth using multiple isotopologues}
\label{sec:multiso} 

\begin{figure}
\center
\begin{tabular}{@{}cc@{}}
\includegraphics[scale=0.45, angle=0]{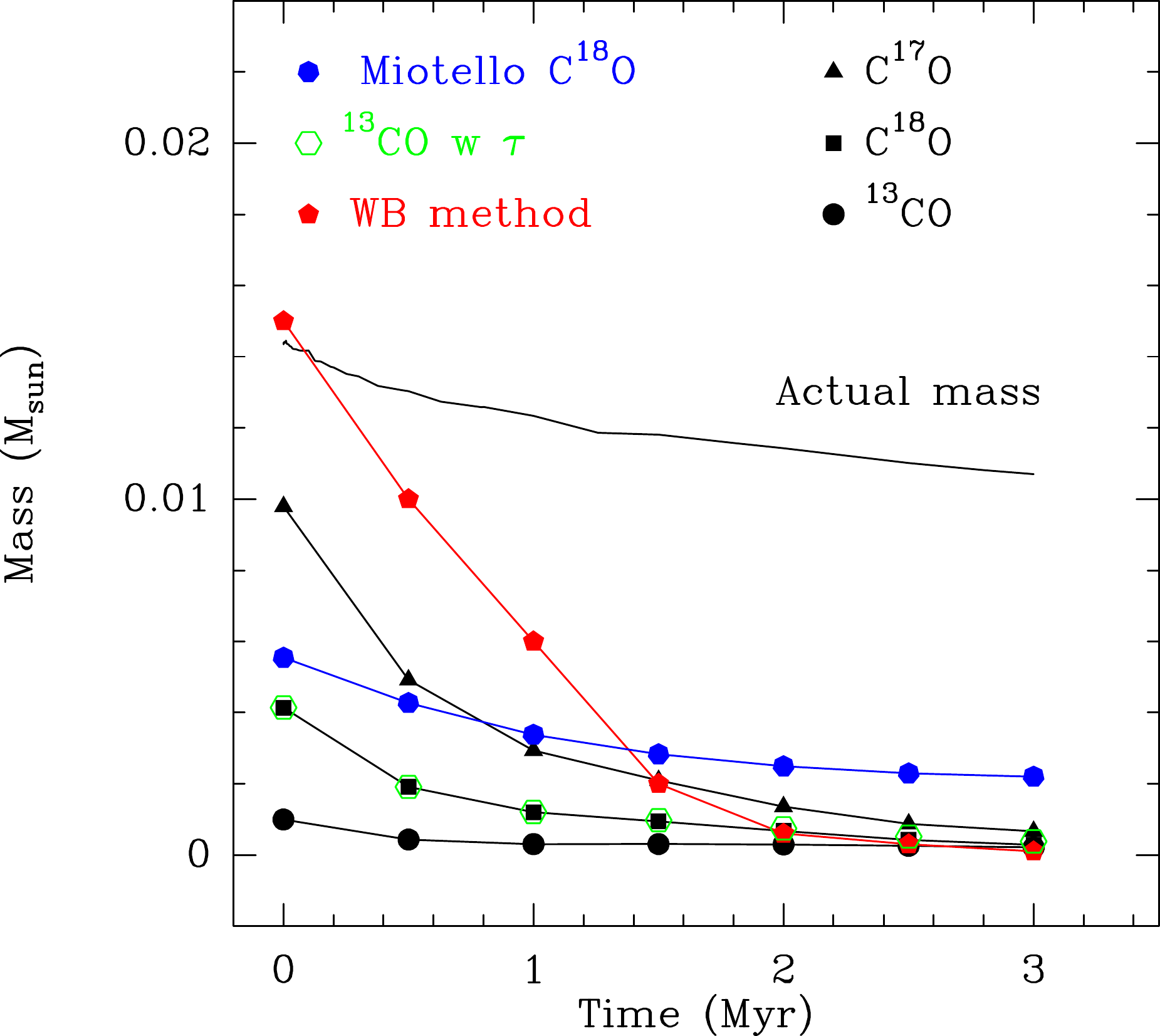} &
\includegraphics[scale=0.45, angle=0]{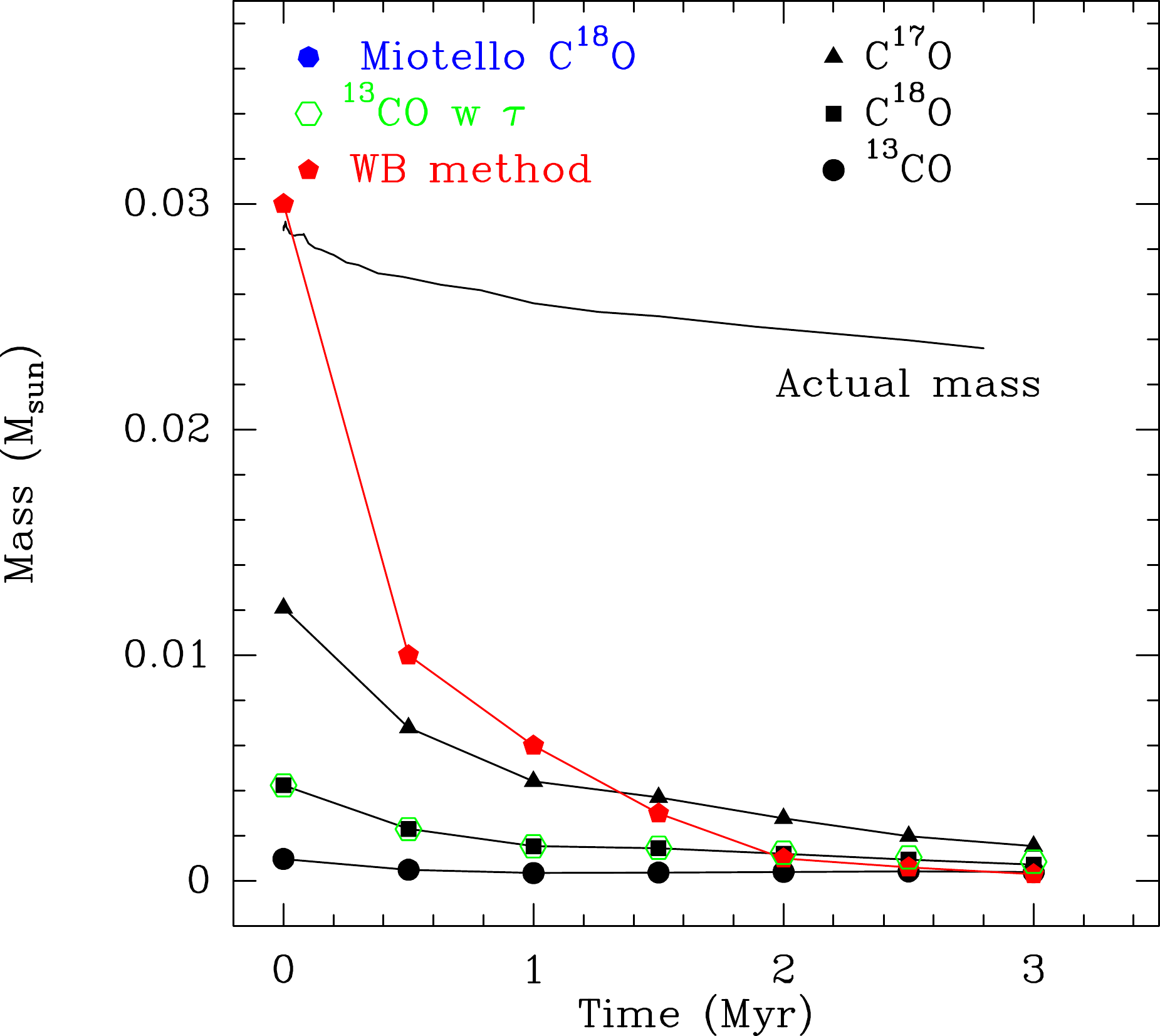} \\
\end{tabular}
 
\caption{
The mass of the disk inferred from the simulated observations is plotted
versus time. The actual mass is shown as a solid line, while the masses
inferred from equation \ref{masseq} and the simulated emission from different
isotopes and using different analysis methods are shown as points. Colored points show mass estimates corrected for optical depth but not \fco, while black points show masses corrected for \fco\ but not optical depth. Labels are explained in the text.
}
\label{massplot}
\end{figure}

One approach to dealing with optical depth is to observe the
same rotational transition in multiple isotopologues
\citep[e.g.,][]{Williams_Best_2014, Miotello_2016}. 
We tried 
a simple correction for our $^{13}$CO-based mass estimate by comparing with C$^{18}$O, using the C$^{18}$O/$^{13}$CO line intensity ratios
plus the initial isotope ratios from the chemical model. 
This correction yielded the hollow green hexagons in Figure \ref{massplot}, which still underestimate disk mass badly, performing only as well as C$^{18}$O observations would on their own.

More sophisticated optical depth corrections can be made with fitting formulae
that relate the \jj21\ line intensity of either $^{13}$CO or
C$^{18}$O (or their ratio) to the total disk mass, based on a suite of models
\citep[e.g.,][]{Williams_Best_2014, Miotello_2016}.
We placed our model line luminosities on Figure \ref{massplot} (left panel) of
the Williams and Best models to estimate mass, producing the red points, which have no correction for \fco. They do well for early times when the CO is more uniformly distributed over the disk, but underestimate the mass badly at later times when \fco\ drops  especially at larger radii.
Most of our points lie near the top of their distribution 
of models, and interpolation between mass models is uncertain by a factor of three.
Second, we used the formulae in equation 2 with coefficients in Table
A.1 of \citet{Miotello_2016} to estimate disk mass from luminosities
of the \jj21\ lines of \coo\ and \cooo. Those from \coo\ underestimated
the mass badly, but those from \cooo\ did better (blue octagons).
In these comparisons, we adjusted their mass estimates to be consistent 
with our assumption
about the atomic carbon abundance, but not for the fraction of carbon in CO.
In a recent paper, \citet{Miotello_2017} find the same result;
applying their models to the Lupus data \citep{Ansdell_2016_ALMA_Lupus}
data leads to very low gas to dust ratios (less than 10 in most cases)
unless CO is depleted. The Miotello et al.\ grid of chemical models was optimized to treat CO self-shielding and isotopic fractionation, which our chemical models also include; the low gas/dust ratios implied by C$^{18}$O and $^{13}$CO observations must result from other CO-depletion pathways.

Clearly, both optical depth and CO chemical depletion must be accounted for when measuring disk masses: Figure \ref{massplot} shows that measurements incorporating one correction, but not the other, will fail.
For masses based on a single emission line, so that no optical depth correction is possible, the isotopotologue that performs the best is C$^{17}$O, which
underestimates the mass by a factor of 2-3 if the correct value of \fco\ is known (which it will not be in real astrophysical situations).
In Appendix \ref{sec:rotdiag}, 
we show that the \jj10\ line does much better than the
more commonly used \jj21\ or \jj32\ lines. 
We believe that this is primarily due
to its lower optical depth and lesser sensitivity to temperature.

The combination of the models of \citet{Miotello_2016} and our calculations
of \fco\ would improve the mass estimates. For example, if we use our ``insider information'' (i.e. model-derived knowledge) about \fco\ as a function of time for our 0.015 \msun\ disk to further correct the masses
from the formulae in \citet{Miotello_2016}, the resulting masses are
accurate to within a factor of two at all ages.
Unfortunately, to apply this method to observations, one needs to know the
disk age. We explore the issues arising from that fact in the next section.

\section{The Age-Mass Degeneracy}
\label{sec:agemass}

\begin{figure}[tbh]
\begin{center}
\begin{tabular}{@{}cc@{}}
 \includegraphics[width=0.45\textwidth]{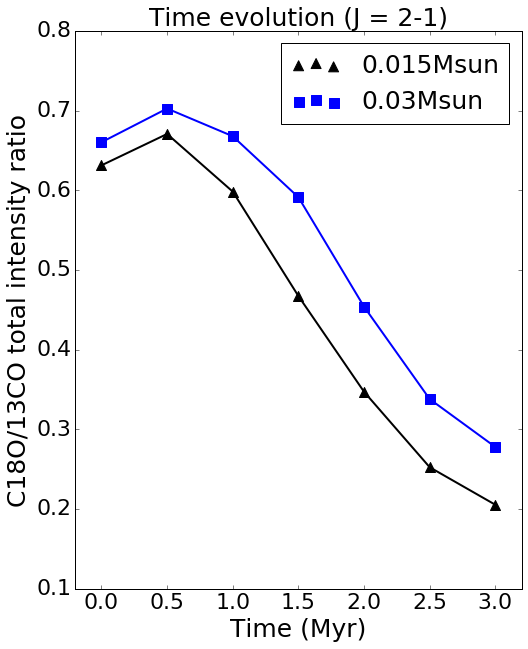} &
  \includegraphics[width=0.45\textwidth]{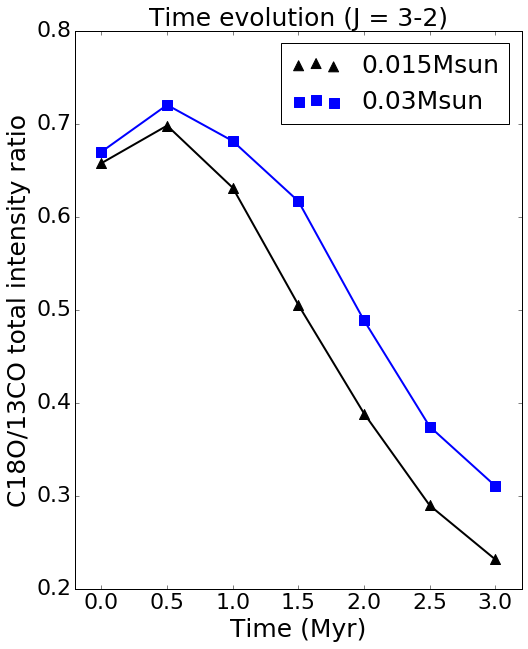} \\
 \end{tabular}
\caption{Ratio of total intensities of C$^{18}$O and $^{13}$CO lines from for the $0.015 \Msun$ and the $0.03 \Msun$disks. We show the results for the J=$2 - 1$ lines on the left, and those for J = $3-2$ lines on the right.}
\label{fig: time_line_ratios}
\end{center}
\end{figure}

The time dependence of the fraction of C in CO (see Fig. \ref{fcovst})
implies that one needs to know the disk age to apply a mass correction factor.
Ages of host stars are not generally known to better than $\pm 1$ Myr.
Worse yet, the ``age" in our models is in some sense a chemical age 
because the speed of chemical evolution will depend on the ionization
sources. EXor outbursts and X-ray flares may also introduce brief periods of intense ionization \citep[e.g.][]{audard14}. If we don't know the chemical age, there is
  an age-mass degeneracy---a massive, old disk
may look similar to a young, less massive one in CO rare isotopologue
emission. 
Figure~\ref{fig: time_line_ratios} shows an example of the age-mass degeneracy.  The more massive disk reaches the same C$^{18}$O/$^{13}$CO total intensity ratio about 0.5 Myr later than the less massive disk for both \jj32\ and \jj21.

CO-based mass measurements are further complicated by the fact that the depletion occurs in the outer disk, where most of the mass resides. The decline in CO isotopologue line intensity as a function of time is precipitous: in the $0.015 \Msun$ model, the intensity of $^{13}$CO J=3-2 emission drops from 
$2.72 \jykms$ by $82\%$ to $0.48 \jykms$ over the $3$~Myr disk evolution, and 
the intensity of C$^{17}$O J=3-2 drops by $95\%$ from $1.15\jykms$ to $0.06 \jykms$ (Fig. \ref{fig: time_evolution_intensities}). 
For comparison, the total disk mass within 70~AU of the star drops by about 12\% over 3~Myr and \fco\ drops by a factor of 2-3. Clearly, a simple
correction for \fco\ does not capture all of the decrease in line
emission with time.
Without a good model of CO abundance as a function of radius and 
time, line-intensity measurements yield disk mass estimates that are accurate 
to factor of two at best and are systematically underestimated.



\section{Diagnosing CO chemical depletion }
\label{sec:results_agemass}

In section \ref{sec: DiskModel} we demonstrated that the CO abundance decreases with both radius and time. In Section \ref{sec:measure_mass}, we showed
that other effects add to the underestimation of disk mass and that the
underestimate gets worse with age. Given
that star ages can be uncertain by over 1~Myr
\citep[e.g.,][]{soderblom14}---the timescale over which we observe CO
depletion in our model---interpreting CO observations requires a more
direct CO chemical depletion  indicator than the star age. The need for a
disk-based CO chemical depletion  indicator is especially evident given the
likelihood that disks with different masses or incident cosmic-ray
fluxes may evolve at different speeds.

We explore here three types of observations that can diagnose CO chemical depletion . First, isotopologue intensity ratios for a transition change with time, revealing departures from the ISM CO/H$_2$ ratio. Second, line profile
shapes for the most optically thin isotopologues, C$^{17}$O and C$^{18}$O, widen over time as the outer disk loses CO, while $^{13}$CO and CO profile
shapes evolve very little.  Third, spatially resolved observations will reveal
CO chemical depletion  patterns. While using intensity ratios to diagnose CO chemical depletion  requires less observing time, high signal-to-noise line profiles or
spatially resolved observations contain valuable information that could allow observers to reconstruct the radial distribution of CO gas. We present all three strategies here, beginning with intensity ratios.

\subsection{How intensity ratios reveal CO chemical depletion }
\label{sec: lineratios}

The relative strengths of emission lines for different isotopologues change significantly over time. This is because different isotopologues react to the change in CO abundance differently due to their various optical depths. 
Re-examining Figure \ref{fig: time_line_ratios}, we see that after $1\Myr$, the \cooo/\coo\ intensity ratios decrease steadily as 
the disk-averaged CO abundance declines monotonically (Fig. \ref{fcovst}). One could use Figure \ref{fig: time_line_ratios} to 
estimate the
chemical age of an observed disk. However, the predicted intensity ratios are higher for the more massive
disk, so there is still a 0.5~Myr uncertainty in age, which propagates into a mass uncertainty.


Intensity ratios are the most rudimentary diagnostic of CO chemical depletion. \citet{Williams_Best_2014}, \citet{Miotello_2014}, and \citet{Miotello_2016} show how isotope-selective photodissociation, varying disk radii, and CO freezeout (or lack thereof) can affect intensity ratios, leading to a wide dispersion of C$^{18}$O/$^{13}$CO for disks of the same mass.

\subsection{How line profiles reveal CO chemical depletion }
\label{sec: lineprofiles}

The line profiles contain information not available in the integrated
intensities. Here we demonstrate how a comparison of normalized 
line profiles from $^{13}$CO and C$^{18}$O or C$^{17}$O can
diagnose CO chemical depletion, even over a factor of two range in disk mass.
Figure \ref{fig: line_evo} shows that the normalized 
line profiles of the rarer isotopologues become broader than those of
the more common isotopologues as time proceeds.
Since CO chemical depletion  happens primarily in the outer part of
the disk where Keplerian velocities are small, the fraction of radiation
in the high-velocity line wings increases with time for optically thin
lines. Optically thick lines can mask CO chemical depletion: 
the CO column density has to first decline to the $\tau \approx 1$ threshold
(where $\tau$ is the optical depth in the line center) before 
intensity changes start to track abundance changes. 
For the $0.015 \Msun$ disk, the optically thin C$^{17}$O 
and C$^{18}$O emission tracks the changes in CO column densities well, 
so that the line profile gets significantly broader after $1$ Myr. 
$^{13}$CO ($J = 3 \rightarrow 2$) emission is optically thick well beyond 
the CO chemical depletion  
radius of $\sim 20 \AU$, so does not reveal reductions in CO column density 
with time. Similar trends are found for the $0.03 \Msun$ disk.  Because of the 
relatively high column densities, the change in C$^{18}$O line profile is not 
obvious until $3\Myr$, but C$^{17}$O still tracks the reductions in CO column
density very well. Clearly, line
profiles of rare isotopologues can diagnose CO chemical depletion .

However, emission line profiles reflect not only the CO/H$_2$ abundance
ratio, but the temperature and density structure of the disk as well.
The temperature of our model disk decreases with time due to the young star's
dimming as it moves down the Hayashi track, and the density structure
changes as the disk viscously evolves. To isolate the effect of CO
depletion from density and temperature effects, we set up control models with a
constant CO/H$_2$ ratio throughout the disk and compare the emission
line profiles produced by these with those produced by our fiducial,
CO-depleted models. In the constant CO/H$_2$ models, all atomic carbon
available for gas-phase reactions is assumed to be in CO, and relative
abundances of CO isotopologues are determined only by the abundance
ratios of the isotopes (see Table 2 of Paper 1).  The abundances
normalized to the total proton density are $7.21\times 10^{-5}$ for CO,
$9.34\times 10^{-7}$ for $^{13}$CO, $1.44\times 10^{-7}$ for C$^{18}$O,
and $3.13\times 10^{-8}$ for C$^{17}$O. 

\begin{figure*}[ht]
\centering
 \includegraphics[width=0.7\textwidth]{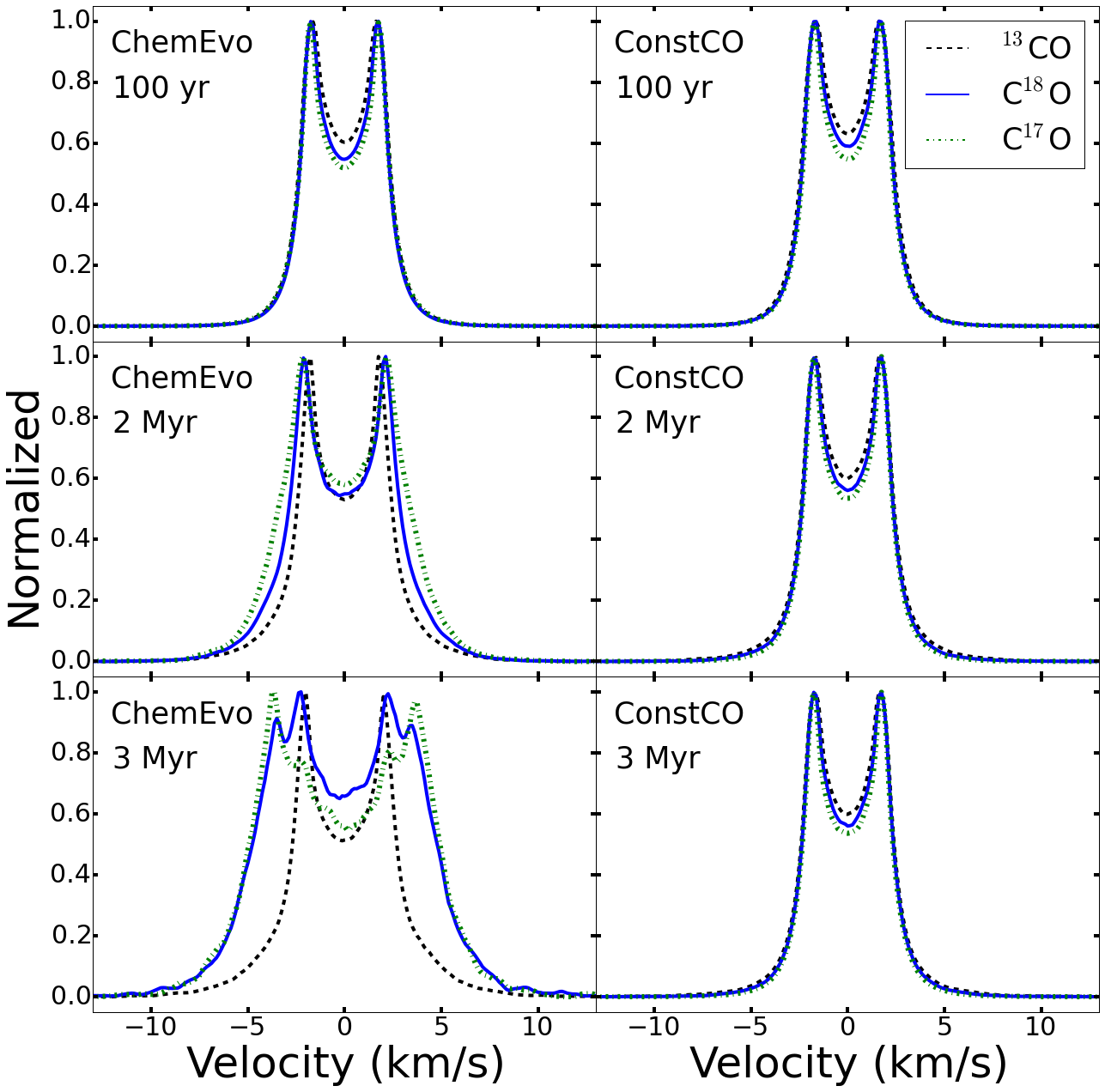} 
  \caption{A comparison of normalized emission line profiles for various
  isotopologues are shown for the chemical evolution model and the
  constant CO model. In the constant CO model, all carbon available for gas phase
  reactions is assumed to be in CO, and abundances of CO isotopologues
  are assumed to be determined by atomic abundances of the isotopes.
}
\label{constco}
\end{figure*}

Figure \ref{constco} shows side-by-side normalized line profiles for the 
$0.015$ \msun\ disk
with evolving chemistry and the one with constant \fco. By 2 Myr, the
profiles of rare isotopologues are noticeably wider than those of more
common isotopologues and by 3 Myr, they are quite distinctive. These
line profiles comparisons would provide clear evidence for ongoing
CO chemical depletion and a potential way to correct for it (though there may be multiple rings of gas in some disks, which would complicate the line profile analysis; see \citealt{cleeves16}). However, the actual (un-normalized) lines are very weak (Fig. \ref{fig: line_evo}), 
so diagnosing CO chemical depletion by comparing line profiles would be an expensive method in terms of observing time. Furthermore, variation in turbulent speeds between different layers of the disk could also produce different line profiles for $^{13}$CO, C$^{18}$O, and C$^{17}$O \citep{flaherty15, simon15}, an effect we have not explored here.

\subsection{Spatial distributions of CO isotopologues}
\label{sec: spatial}
\begin{figure*}[ht]
\centering
\begin{tabular}{@{}cc@{}}
 \includegraphics[width=0.4\textwidth]{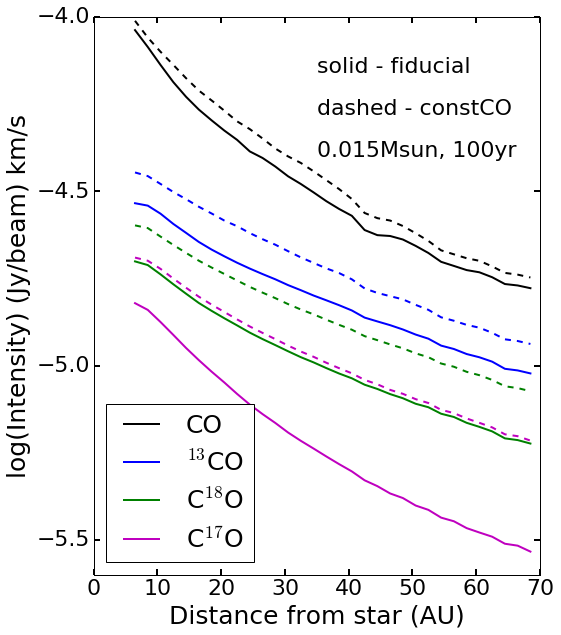} &
 \includegraphics[width=0.4\textwidth]{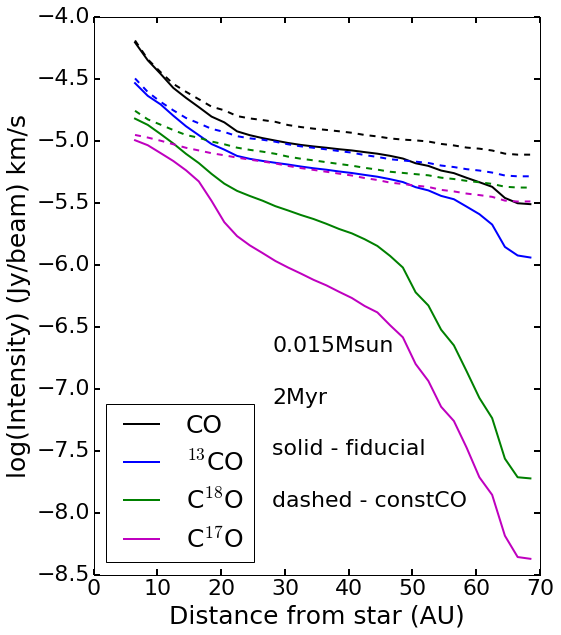} \\
  \includegraphics[width=0.4\textwidth]{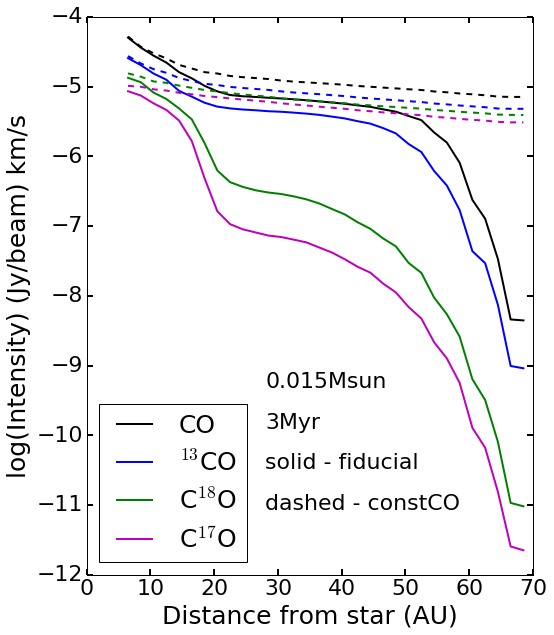} &
 \includegraphics[width=0.4\textwidth]{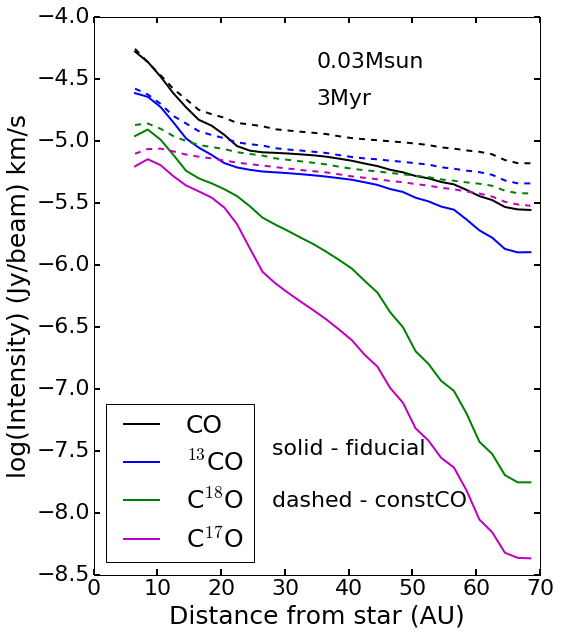} \\
  \end{tabular}
\caption{ Logarithm of the azimuthally averaged, velocity-integrated intensities of the \jj21\ line as a function of radius. For a 10 km baseline, ALMA has a 0\farcs03 resolution for the \jj21 line.}
\label{fig: spatial_comparison}
\end{figure*}

Our chemical evolution models also predict a characteristic radial dependence of the CO abundance. To translate this dependence into observables, we average the velocity-integrated, continuum-subtracted emission from our model disks in rings of $2$ AU in radius, and compare the spatial distributions of the \jj21\ emission line from the fiducial models to that from the constant CO models in Figure \ref{fig: spatial_comparison}. The models shown in Figure \ref{fig: spatial_comparison} were run in face-on geometry for simplicity; all other simulated emission in the paper is calculated for $30^{\circ}$ inclination. The optical depth effects are enhanced in the face-on disks because the lines are not spread out by rotation. If a disk inclination angle is known, models could be run for that situation.

Results from the fiducial $0.015 \Msun$ disk are presented in the upper-left (early time), upper-right (2~Myr), and lower-left (3~Myr) panels. At the beginning of the evolution, the integrated intensities are lower in the fiducial model but the profiles are very similar to those in the constant CO disk for all isotopes. This is because part of the carbon is locked in CO$_{2}$ in the fiducial model, whereas the constant CO model has all available carbon in CO. As the disk evolves, the intensities beyond 20 AU decrease dramatically in the fiducial model due to the depletion of CO, and the differences between the fiducial model and the constant CO model are the greatest for C$^{17}$O lines, which have the lowest optical depth. We see sharp drops of intensity in models with $0.03 \Msun$ (lower right) as well. However because the chemical depletion happens more slowly in the $0.03 \Msun$ disk, the intensity profiles at 3 Myr of the $0.03 \Msun$ disk resemble those of the $0.015 \Msun$ disk at 2 Myr. 

\begin{figure*}[ht]
\centering
\begin{tabular}{@{}cc@{}}
\includegraphics[width=0.4\textwidth]{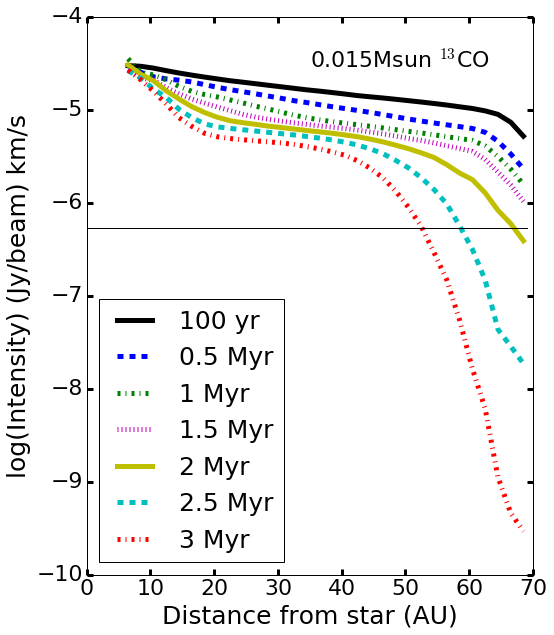} &
\includegraphics[width=0.4\textwidth]{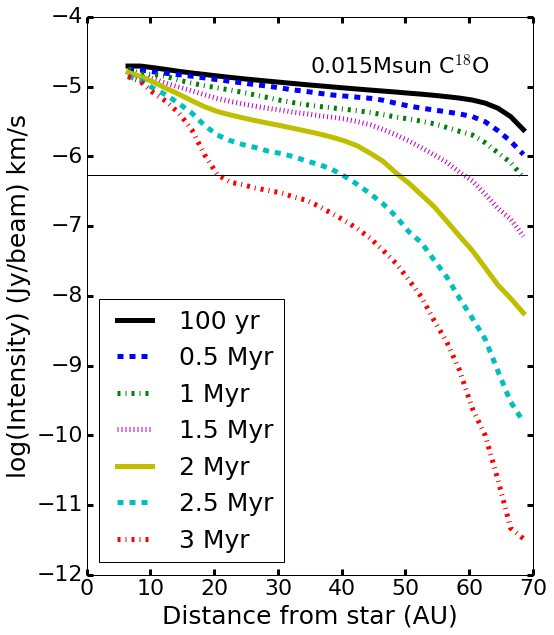} \\
 \end{tabular}
\caption{Time evolution of $^{13}$CO and C$^{18}$O $\jj21$ lines for the $0.015 \Msun$ model at 30 degree inclination. All other parameters are the same as the above figure.}
\label{fig: spatial_time}
\end{figure*}

 The effect of age is detectable with ALMA given enough integration time. We plot the time evolution of $^{13}$CO and C$^{18}$O lines for the $0.015 \msun$ disk in Figure \ref{fig: spatial_time}, and the detection limit as a horizontal line at log $= -6.27$.  At  resolution, we get to 540 mJy in 1 km$/$s resolution in 10 hours. At $0\farcs03$ spatial resolution, with a velocity channel of width 1 km$/$s , we reach a sensitivity of 540 mJy per beam in 10 hours. The line width for a perfectly face-on disk could be small, but it only takes a small inclination angle to significantly broaden the lines (for example, it takes only 6 degrees for Keplerian velocity to broaden the line to 1 km/s at 10 AU around a solar mass star). $0\farcs03$ will resolve 4.2 AU at 140 AU and matches our resolution in the figure. For the purpose of illustration, we use the $\jj21$ lines from a disk of $30^{\circ}$ inclination. The apparent edge of the disk in $^{13}$CO moves from beyond 70 AU to about 55 AU, and from beyond 70 AU to about 20 AU in C$^{18}$O emission, well within the condensation front of CO in both cases. The migration of the ``fake'' snowline is a combined effect of the drop of CO emission intensities from the outer disk and the detection limit of the observation.

In disks where spatially resolved imaging is possible, a comparison between the C$^{18}$O and the CO or $^{13}$CO spatial distribution should reveal CO chemical depletion, and may even provide enough information to re-construct the CO distribution with the help of a chemical model. Finally, after 3 Myr of evolution, we see that two rings of C$_2$H have formed in the surface layers of the disk: a narrow ring at $1-3$ AU and a broader ring at $10-20$ AU. \citet{Bergin_2016} found that C$_2$H formation is possible only with a strong UV field and C/O$ > 1$, conditions which are replicated in our model disk surface layers due in part to the breakdown of CO molecules. We will explore the detectability of the C$_2$H rings and the degree to which they may indicate CO depletion in future work.

\section{Consequences for Observations}
\label{sec: obs}

Our main result is that current observations and interpretations of CO
isotopologues toward disks around solar-mass stars likely underestimate gas masses by substantial
amounts. The surveys so far have emphasized
shallow observations of many disks with modest spatial resolution.
For example,
\citet{Ansdell_2016_ALMA_Lupus} surveyed $89$ protoplanetary
disks in Lupus with ALMA and found very low gas masses when assuming
an ISM CO abundance. The analysis of those
observations has led to the conclusion that almost no disks have masses
that exceed the MMSN. This result would appear to conflict with the
high fraction of stars with evidence for planetary systems.

Our predicted line intensities are weak, and deriving correct masses
requires deeper integrations to obtain either line profiles of weak lines or
very high spatial resolution. Until those observations are available,
we can offer only rough estimates of how much more massive the disks
may be. As an example, we ask the following question: 
if we apply a simple correction for \fco, how many of 
the disks in Table 3 of \citet{Ansdell_2016_ALMA_Lupus} might contain
the mass of $0.01$ \msun, the MMSN. If the mean age of stars in Lupus
is $3\pm2$ Myr, we should use $\fco = 0.136$, the value at 3 Myr for the
0.015 \msun disk, the closest to a MMSN. After applying this correction 
factor, 7 of 36 best guess masses exceed the MMSN, and 26 of 36 maximum
masses exceed the MMSN, where best guess and maximum masses are defined by Ansdell et al. Within the uncertainties, a significant fraction of the disks
in Lupus could contain enough mass to make a planetary system like ours.
Recent analysis of similar observations toward disks in Cha I indicate
a deficit of gas mass similar to, or worse than, that seen in the Lupus
disks (Feng et al. 2017). For an age of 2 Myr and the 0.015 \msun\ disk,
the value of \fco\ is 0.18, which could increase the masses by a factor
of about 5, leaving almost all disks in Cha I still short of the MMSN. 
Almost all these stars are lower in mass than our model star; models
tuned to these stellar parameters would be needed to draw further 
conclusions.

CO chemical depletion  also affects estimates of gas/dust mass
ratios.  Assuming an ISM-like CO/H$_2$ abundance ratio, \citet{Ansdell_2016_ALMA_Lupus} found a wide range of gas/dust ratios, 
with a median ratio of $\sim 15$, which is much smaller than the canonical value of $100$ observed in the ISM. 
\citet{Miotello_2017} have shown that this problem persists when
their models with radiative transfer and isotope-selective
photodissociation are used.

\begin{figure*}[ht]
\centering
 \includegraphics[width=0.7\textwidth]{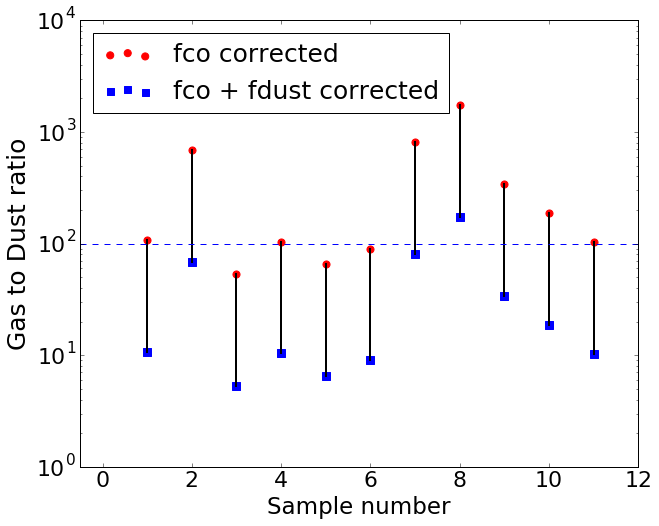} 
  \caption{Corrected gas-to-dust ratios of the 11 stars in 
\citet{Ansdell_2016_ALMA_Lupus} with both $^{13}$CO and 
C$^{18}$O detections. Red dots are values corrected for the CO 
fraction of 0.138, and blue squares are values corrected for both the 
CO fraction and the fraction of solids in dust ($10\%$). The stars 
are in the order as presented in \citet{Ansdell_2016_ALMA_Lupus} 
(shown as the sample number in Fig.~\ref{fig: ansdell_g2d_corrected}) 
and the ordering has no particular meaning. The blue dashed line 
shows the ISM value of 100. The ISM value falls into the 
possible range of gas to dust ratio for 7 out of 11 stars.}
\label{fig: ansdell_g2d_corrected}
\end{figure*}

We argue that the low gas-to-dust ratios measured in \citet{Ansdell_2016_ALMA_Lupus} 
are likely a result of a low CO/H$_{2}$ ratio due to CO chemical depletion. In our 0.015 \msun disk model, the closest 
to a MMSN, only 13.6$\%$ of carbon is contained in CO gas by 3 Myr. If the 
mean age of stars in Lupus is $3\pm2$ Myr, and we apply the simple correction 
for CO fraction (\fco = 0.136) to the Lupus data with both 
$^{13}$CO and C$^{18}$O detections, we obtain gas-to-dust ratios of 
53 to 1700, with a median gas-to-dust ratio 
of 108. The gas-to-dust ratios correcting for the fraction of C in CO in our models 
are plotted in Fig. ~\ref{fig: ansdell_g2d_corrected}.

The high values of the gas-to-dust ratios in the $\fco$ corrected gas-to-dust 
ratios could be a combined result from grain growth, weak-ionizing environments, 
and young disk ages. 
Gas/dust mass ratios are not necessarily the same as gas/{\it solid}
mass ratios: the apparent dust mass is typically calculated from sub-mm
dust emission, which is mostly sensitive to dust grains around sub-mm
size or smaller. Once dust grains grow into cm-size pebbles, the reduction of
observable dust mass could drive the gas/dust mass ratio to {\it larger} than
the canonical molecular cloud value of $100$. If only $10\%$ of the solid 
mass is in the form of dust observable in $890$ \micron, 
as we assume in our models, one would have to correct 
for the ``dust fraction" ($\fdust$) when measuring the gas-to-dust ratios. 
We apply the correction for dust fraction on the $\fco$ corrected data, 
and plot the gas-to-dust ratios as blue squares in Fig. ~\ref{fig: ansdell_g2d_corrected}. 
Most the disks with high gas-to-dust ratios from the $\fco$ corrections can be come consistent with the ISM value of 100 if additional corrections for grain growth are included.

Instead of assuming that dust evolution or disk clearing drives
the variation in observed Lupus gas/dust ratios, we point to the large
age spread of pre-main-sequence stars in Lupus. Ansdell
et al.\ quote 1-3~Myr, while \citet{mortier11} find either 0.1 to
$>15$~Myr or 0.3 to $>15$~Myr depending on the model isochrones (though
the $>10$~Myr objects were not detected by Ansdell et al.\ and are
therefore not represented in Figure \ref{fig: ansdell_g2d_corrected}). The large spread of
gas/dust ratios calculated by Ansdell et al.\ is likely the result of
the variation of CO/H$_2$ abundance ratio with protostellar age.
Given the uncertainties in both gas and solid mass,
the data so far do not rule out ISM gas/dust ratios.

There are a few caveats to bear in mind regarding these simple mass correction
factors.
First, real disks evolve at different rates: a disk with a high
incident cosmic-ray flux loses CO more quickly than a disk shielded by a
``T-Tauriosphere'' that deflects cosmic rays \citep{Cleeves_2013_CR,
cleeves15}. Even if star ages could be measured perfectly, the ages
alone do not give enough information about the chemical evolutionary
stage of the disk to compute a CO/H$_2$ ratio. In other words, our line 
profile diagnostics are measuring a ``chemical age.'' Comparison of $^{13}$CO
and C$^{18}$O line profiles gives an empirical diagnostic of the level
of CO chemical depletion .

Second, chemical depletion of CO operates on a million-year
timescale. If episodic
accretion of the type observed in FU~Orionis outbursts happens during
the T-Tauri phase at intervals smaller than $\sim 1 \Myr$
\citep[e.g.,][]{dunham10, kim12, martin12, green13}, desorption of CO$_2$
ice followed by CO$_2$ dissociation from cosmic ray-induced photons could raise the
CO gas abundance throughout the disk, not just in the inner 20~AU as in our current model disk. \citet{cieza16} have already proposed
chemical alteration in the disk surrounding V883~Ori, which is now
mid-outburst. We have not modeled FU Orionis outbursts---our model
T-Tauri star evolves smoothly on the Hayashi track.

Other processes may further deplete CO in the outer regions of disks. For example, \citet{Xu_Bai_2017} describe a process of ``runaway freeze-out", which considers vertical transport at fixed radii. They find that higher layers, too warm for freeze-out, can become depleted by transport to the colder, lower layers. \citet{kama16} suggest the same mechanism to explain the low atomic C and O abundances in the TW~Hya disk. Applied to CO, the vertical transport/freezeout effect would primarily act at larger radii than we consider and affect primarily the emission from the more common, hence more optically thick, isotopes. Applied to the complex organics that sequester the carbon in our models, runaway freezeout could remove even more carbon from the gas phase, allowing for even higher H$_2$ masses to be consistent with observed gas-phase carbon abundances.

Our model's growing CO abundance in the inner disk and CO chemical depletion in the outer disk would be observationally almost identical to a disk with CO frozen onto grain surfaces in the outer disk, followed by inward radial drift of the grains and desorption of CO in the warm inner disk. The radial-transport effect, first applied to water ice, has been suggested on theoretical \citep{Ciesla_Cuzzi_2006, Du_2015} and observational \citep{Hogerheijde_2011, Kama_2016} grounds, diagnosed from the presence of rings of small hydrocarbon chains \citep{Bergin_2016} and ammonia gas \citep{Salinas_2016} in the outer disk of TW Hya, and inferred in cases where C/O$ >1$ \citep{Bergin_2016}. Robust temperature measurements in the outer disk would be required to distinguish between chemical CO chemical depletion and CO freezeout followed by grain inspiral.

\section{Conclusions}
\label{sec:conclusion}

Our key findings and suggestions for observing strategies are summarized below. 

\begin{enumerate}
\item CO abundance varies both with distance from the star and as a function of time as CO is dissociated and the carbon gets sequestered in organic molecules that freeze onto grain surfaces (chemical depletion) on a million year time scale (\S \ref{sec: DiskModel}). CO chemical depletion  will cause very large underestimates in gas mass and gas-to-dust ratios when CO observations are used. One would need to correct for the chemical depletion of CO in order to correctly estimate the disk gas mass. 
The CO abundance correction factor ranges from 3 to 8 for the models we have run
(\S \ref{sec:measure_mass}).

\item Even though CO is destroyed by ionized helium throughout most of the disk, it has a higher-than-interstellar abundance in the inner 20 AU of the disk, where the temperature is relatively high. Adopting a constant $T = 20$~K for the CO reservoir will underestimate
gas masses by a further factor of about 2 (\S \ref{sec:measure_mass}).

\item The high CO abundance in the inner disk also results in high optical depth even for the mostly optically thin isotopologue we investigated (C$^{17}$O), and one could underestimate the disk gas mass due to the optical depth effect even after correcting for the CO chemical depletion in the outer disk and using the correct CO-averaged temperature (\S \ref{sec:measure_mass}).

\item The CO-abundance time evolution also introduces a disk age-mass degeneracy---a massive, old disk may look similar to a young, less massive one in CO rare isotopologue emission (\S \ref{sec:agemass}). 

\item One can diagnose CO chemical depletion by comparing the line intensity ratios (\S \ref{sec: lineratios}) or emission line profiles of multiple isotopologues (\S \ref{sec: lineprofiles}). If the disk is spatially resolved, one can also use spatial distribution of CO beyond $20$~AU to probe the chemical depletion of CO. 
(\S \ref{sec: spatial}).

\item Very different CO optical depths in different parts of the disk produce a complicated rotation diagram that does not probe the disk temperature well. One would underestimate the average temperature of CO molecules by deriving it from the lowest two J values (\S \ref{sec: Testimates}). 

\item Higher-$J$ lines underestimate the disk mass by more than do lower-$J$
lines. The higher-$J$ lines miss low temperature gas.  We suggest using low excitation lines (e.g., \jj10) to estimate the disk mass to minimize the temperature and optical depth effects.
(\S \ref{sec: transitions}).

\item The strategy that comes closest to recovering the correct mass
for our models is to use the formulae from \citet{Miotello_2016} {\it and}
to divide by the \fco\ from our models (\S \ref{sec:measure_mass}). 

\item If we correct the ``best-guess'' gas masses in \citet{Ansdell_2016_ALMA_Lupus} by the
smallest value of \fco\ (the value at 3 Myr for the 0.015 \msun\ disk),
7 of the disks could have masses
of the MMSN; if we apply our \fco\ correction to the maximum masses, 26 of 36 could
reach the MMSN (\S \ref{sec: obs}). The age of Lupus is $3\pm2$ Myr \citep{2014A&A...561A...2A}, so this correction is reasonable. 
A recent survey of disks in Cha I suggest almost no disks with MMSN
masses.
While still problematic for planet formation models, correcting for
CO chemical depletion  suggests better numbers than finding that no disks contain the MMSN. 

\item Given reasonable uncertainties in both gas and solid masses,
many observed disks can have gas/dust ratios consistent with the ISM value of
100 (\S \ref{sec: obs}).

\end{enumerate}

Work by MY, KW, SDR and NJT was supported by NASA grant NNX10AH28G
and further work by MY and SDR was supported by NSF grant 1055910.
This work was performed in part at the Jet Propulsion Laboratory, California
Institute of Technology. NJT was supported by grant 13-OSS13-0114 from
the NASA Origins of Solar Systems program. 
MY was supported by a Continuing Fellowship from the University of
Texas at Austin. 
We acknowledge helpful input from L. Cleeves, E. Bergin, E. F. van Dishoeck, K. \"{O}berg, J. Huang, M. Ansdell, and A. Kraus. We are grateful for Feng Long et al.\ sharing their paper in advance of publication, and Huang et al.\ sharing processed ALMA data for our model comparison. 

\software{RADMC (http://www.ita.uni-heidelberg.de/ dullemond/software/radmc-3d/), LIME \citep{Brinch_Hogerheijde_LIME_2010}}

\addcontentsline{toc}{chapter}{Bibliography} 
\bibliographystyle{aasjournal}
\bibliography{references,nje}

\appendix
\section{Sensitivity to model parameters}
\label{app: modelpars}

The effect of our assumption of LTE in the CO level populations and the
fixed vertical grid are explored in the next two sub-sections.

\subsection{LTE vs. NLTE}
\label{app: NLTE}


LIME is capable of computing energy level populations either in local
thermodynamic equilibrium (LTE) or in the more complex non-LTE case,
where the kinetic temperature and excitation temperature are different.
In Paper 1, the lowest density found in the modeled region within 70~AU
of the star is about $10^9$ hydrogen molecules
per cubic centimeter. Even in the most diffuse regions of our model
disk, the density should be high enough for collisions to dominate CO
excitation so that LTE is a good approximation for the energy level
population.

\begin{figure}[tbh]
\begin{center}
\includegraphics[scale=0.4]{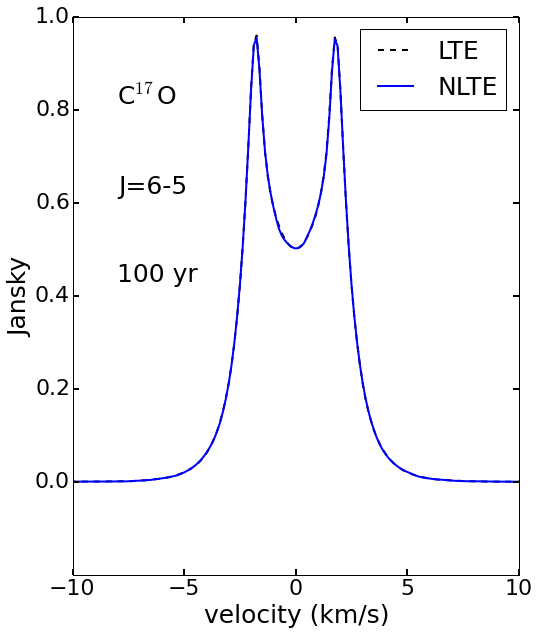}
\caption{C$^{17}$O J=$2-1$ rotational emission at the beginning of the evolution.}
\label{fig: LTE_NLTE}
\end{center}
\end{figure}

We demonstrate that the LTE approximation is appropriate for our model
disk by comparing the results of LTE and non-LTE models for different CO
isotopologues at multiple epochs. For each emission line of each
isotopologue, the line profiles from the two models are
indistinguishable. In figure~\ref{fig: LTE_NLTE} we show the worst case
example of C$^{17}$O J=$6-5$ emission at the beginning of the disk
evolution. The high excitation energy of the
J=$6-5$ transition means the LTE approximation requires a high collision
rate and is hardest to satisfy, yet the differences between the LTE and
non-LTE line profiles are still negligible.

\subsection{Definition of the disk surface}
\label{subsec:surface}

Another parameter that can influence the computed line profiles is the
placement of the disk surface. In the disk models of
\citet{Landry_2013}, on which the chemical models in Paper 1 are based,
the top surface of the computational grid is placed at the layer where
the Rosseland mean optical depth to the disk's own radiation, integrated
downward from infinity, is 0.2. The $\tau = 0.2$ surface is located
between one and two pressure scale heights above the disk midplane,
depending on the distance from the star and time. Since most of the
low-J CO emission comes from the disk interior where the bulk of the
disk mass is located, the height of the grid surface should have minimal
effect on the line profiles. For higher values of J, more of the
emission would come from the warm surface and modeling line profiles
accurately would require the computational grid to extend well into the
tenuous disk atomsphere. Here we assess the effects of grid surface
placement on our model emission line intensities and profiles.

The chemical model in Paper 1 uses a static grid that does not evolve
with time. Grid cells are spaced logarithmically in radius $r$ and
linearly in aspect ratio $z/r$. At the beginning of evolution, which
roughly corresponds to the start of the T-Tauri phase, the disk has a
high scale height due to heating from the luminous protostar. The scale
height decreases throughout the $3\Myr$ evolution due to both
protostellar dimming and viscous dissipation. As the disk flattens, grid
layers with high $z/r$ begin to empty out. To test the effect of surface
placement on the radiative transfer calculation, we construct a new disk
model by artificially removing the top three {\it filled} (not empty)
$z/R$ layers from each time snapshot of our fiducial model from Paper 1.
At six different epochs, we compute line profiles for the
$1\rightarrow0$, $2\rightarrow1$, $3\rightarrow2$, $4\rightarrow3$,
$5\rightarrow4$, and $6\rightarrow5$ transitions for all CO isotopologues using
both the new ``remove-top'' model and the fiducial model. Removing
emitting layers affects the peak intensities of the optically thin
C$^{17}$O emission lines more than any other isotopologue. To verify
that our radiative transfer models capture essentially all of the disk
emission, we demonstrate how the grid surface placement affects
C$^{17}$O.

\begin{figure*}[ht]
\centering
\begin{tabular}{@{}cc@{}}
 \includegraphics[width=0.4\textwidth]{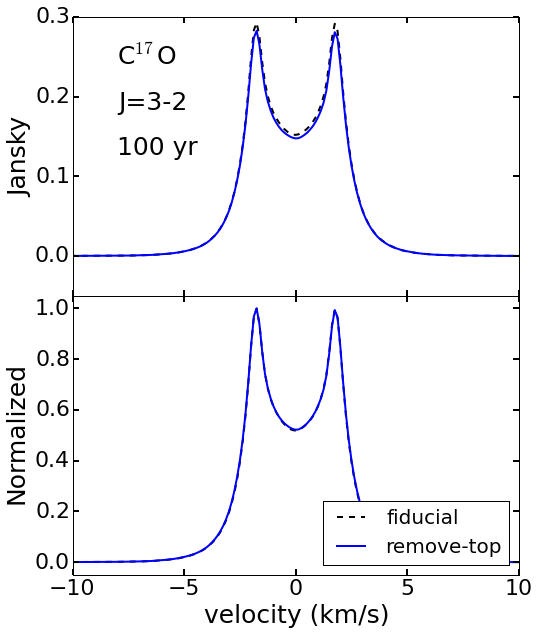} &
 \includegraphics[width=0.4\textwidth]{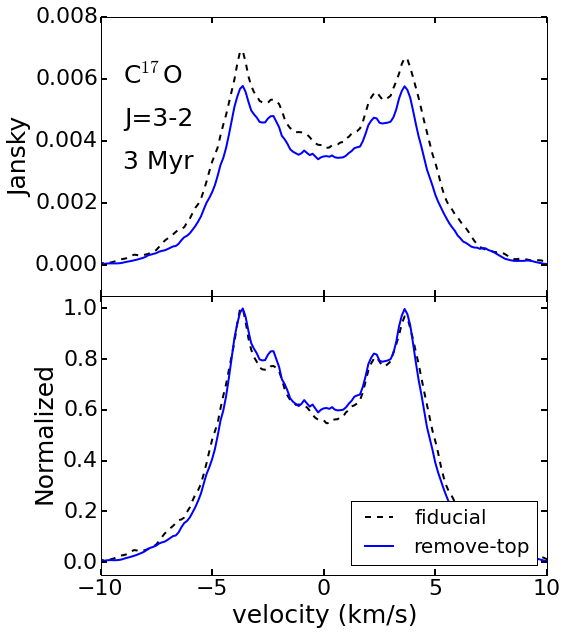} \\
  \end{tabular}
\caption{A comparison of J=$3-2$ emission from C$^{17}$O between the
fiducial models and models with the top-3 existing layers artificially
removed. Left panels: the beginning of the evolution; right panels: the
end of the evolution. The upper panels show the simulated emission lines 
and the lower panels show the line profiles normalized by the total line 
intensities. Removing the top-3 layers of the disk
surface does not change the spectral line profiles at the beginning of
the evolution. However, the mass contained in the surface layers
increases as the disk surface moves closer to the midplane over time, and
line intensities of the surface-removed model are slightly lower across
all frequencies/velocities at the end of the evolution. The effects of
removing the top 3 layers on the line profile remains negligible for low-J
emission.}
\label{fig: fiducial_removetop}
\end{figure*}

At the beginning of disk evolution, the top three grid layers contain
very little mass and have little effect on the emission line intensity
or profile (Figure \ref{fig: fiducial_removetop}, left panel). As the
the disk cools and the top grid layers empty out, the mass contained in
the filled grid layers increases. Line profiles from the fiducial and
remove-top models differ the most after 3~Myr of evolution (Figure
\ref{fig: fiducial_removetop}, right panel).  Yet even at 3~Myr, the
difference in the $J = 3 \rightarrow 2$ peak intensity between the
remove-top and the fiducial model would be difficult to distinguish in
observations, and the normalized line profiles are almost identical.

The differences in peak intensities between the fiducial and remove-top
models become significant for higher-J emission due to the higher
energy needed to populate the upper state. For example, the peak
intensity for $J=6 \rightarrow 5$ differs by roughly $33\%$ between the
fiducial and remove-top models at 3~Myr of evolution. However, lower J 
lines ($J= 1\rightarrow 0$ to $3 \rightarrow 2$) are more commonly used 
for disk studies. In this paper, we focus our line profile discussion on 
\jj32\ and \jj21\ transitions, 
which are the most observationally relevant transitions.  
For our purposes, we have adequately modeled the CO rare isotopologue 
emission, even though our models do not extend vertically to a large number 
of scale heights.

\section{Measuring disk mass from the integrated intensities of CO isotopologues}
\label{sec:rotation_diagram_eqs}

For optically thin emission lines, the total number of molecules per
degenerate sublevel in the upper state can be written simply as
\begin{equation}
\frac{\mathcal{N}_J}{g_J} = \frac{L_J}{g_J A_J h \nu}
\end{equation}
where $L_J$ is the line luminosity, $g_J$ is the degeneracy, $A_J$ is
the Einstein A value for the transition from upper state with quantum
number $J$, $h$ is Planck's constant, and $\nu$ is the frequency of the
transition. $L_J$ is related to the integrated line flux $F_l$ as
\begin{equation}
\label{eq: lj}
L_J = F_l({\rm cgs}) \times 4 \pi D_{\rm cm}^2 =  
1.1964\times 10^{20} \times \fljykms\ (\nu/c)  D_{\rm pc}^2,
\end{equation}
where $D_{\rm cm}$ and $D_{\rm pc}$ are the star's distance in cm and
pc, respectively, and $c$ is the speed of light. For the last part of Eq. \ref{eq: lj} and Eq. \ref{eq: njgj}, F$_{l}$ is in the unit of $ \jykms$, 
 D$_{\rm pc}$ is in the unit of pc, and all the other values are in cgs units.
The Einstein A value for rotational transitions from level $J$ to
level $J-1$ is given by
\begin{equation}
\label{eq: aj}
A_J = \frac{64 \pi^4}{3h} \times (\nu/c)^3 \times |\mu(J,J-1)|^2
= 3.13613\times 10^{-7} \times (\nu/c)^3 \frac{J}{(2J+1)} \mud^2
\end{equation}
where $|\mu(J, J-1)|$ is the electric dipole matrix element and
$\mud$ is the dipole moment measured in Debye ($10^{-18}$ esu-cm).
Combining these, we can write
\begin{equation}
\label{eq: njgj}
\frac{\mathcal{N}_J}{g_J} = 6.468 \times 10^{45} 
\frac{\fljykms \dpc^2}{\bghz^3 J^4 \mud^2}
\end{equation}
where \bghz\ is the rotation constant in GHz. The values of
\bghz\ vary slightly with the isotopologue and are easily obtained
from on-line sources, but for reference, $\bghz \approx 55$ to 58 GHz
for the isotopes discussed here.

\section{Does a single temperature characterize the CO emission?}
\label{sec:rotdiag}

\subsection{Temperature Estimates from Rotation Diagrams}
\label{sec: Testimates}

A temperature estimate is
required in order to evaluate the partition function that will
translate observed intensities into gas column densities. While the disk
temperature can reach $\sim 1400$~K at the dust sublimation front
\citep{muzerolle03}, \citet{andrews05} and \citet{andrews13} suggest
that most submillimeter emission comes from dust at $\sim 20-25$~K and
a fixed temperature of 20 K for the gas is often used in the simplest
methods to measure mass. Here
we will use plots of level populations versus energy of the level above
ground---called rotation diagrams for rotational transitions---to test
whether CO in our model disk is well characterized by a single
temperature. We calculate the rotation diagrams for the $0.015 \Msun$ 
disk and the  $0.03 \Msun$ disk using the equations in Appendix
\ref{sec:rotation_diagram_eqs}, following the method of \citet{green13}. 
 Note that the emission must be optically 
thin (which is often assumed for CO isotopologues) for luminosity to 
be directly proportional to the upper state population.

\begin{figure}
\center
\begin{tabular}{@{}cc@{}}
\includegraphics[scale=0.45, angle=0]{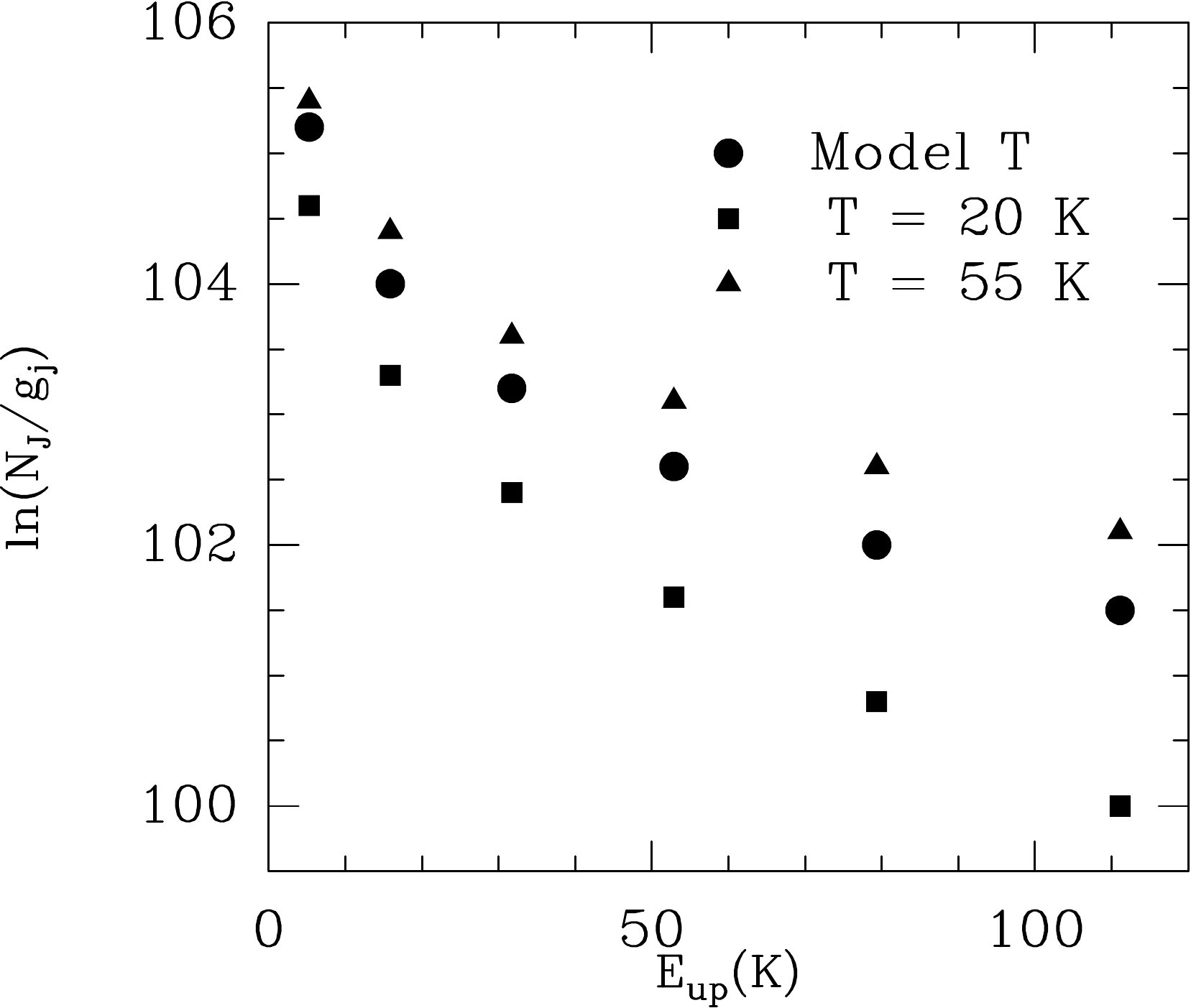}&
\includegraphics[scale=0.45, angle=0]{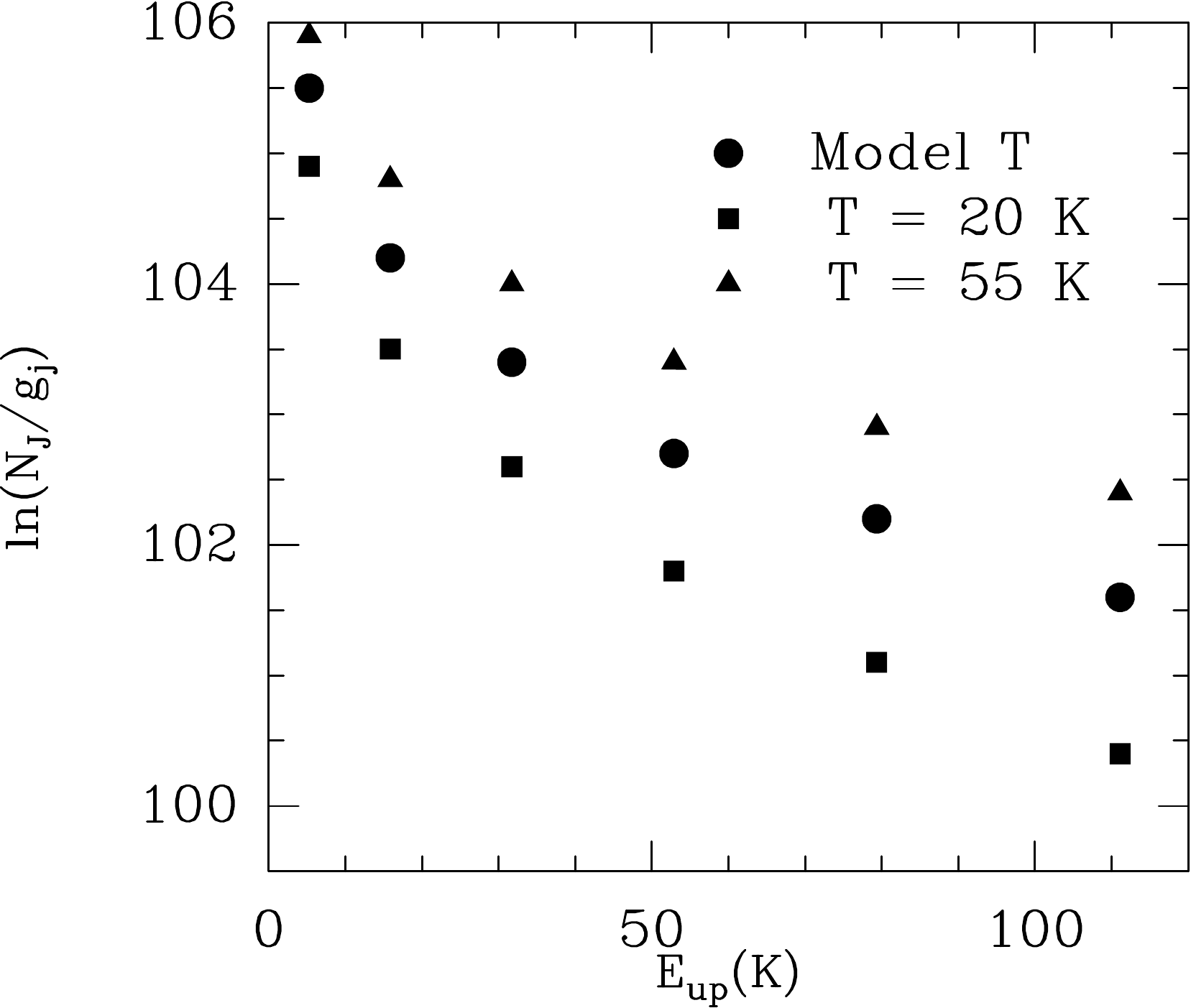}\\
\end{tabular}
\caption{
Rotation diagrams for $^{13}$CO for the disks with mass of $0.015 \Msun$ 
(left) and $0.03 \Msun$ (right) at 2 Myr.
The circles are the values of the number of molecules per sublevel in
the full model. The squares show values for the same model except that
the gas temperature has been fixed at 20 K and the triangles show a model
with $T = 55$ K. }
\label{plotrot2}
\end{figure}

Figure \ref{plotrot2} shows the $^{13}$CO rotation diagram of our $0.015
\Msun$ model disk (left, circles) and our $0.03 \Msun$ model disk
(right), both after 2~Myr of evolution. A gas reservoir with a single
temperature would yield a straight line in $\ln (N_J/g_J)$ as a function
of $E_{\rm up}$. 
Unfortunately, Figure \ref{plotrot2} shows that no single temperature characterizes
the CO level populations, consistent with {\it Herschel} observations of
Herbig Ae/Be and T-Tauri disks by \citet{meeus13}, \citet{Fedele_2013},
\citet{vanderwiel14} and \citet{Fedele_2016}. Even when we construct an artificial disk with
the same density and abundance structure as our 2~Myr
models, but with a constant temperature enforced at all points
 (Figure \ref{plotrot2}, squares and triangles), the rotation diagram
still appears to come from a disk with a range of temperatures because
of the varying optical depth of different transitions. If we
derive a temperature from the lowest two J values, the temperature is
quite low---about 10~K. CO rotation
diagrams, even for rare isotopologues, are not reliable ways to measure
protostellar disk temperatures.

\subsection{Inferred Mass Depends on the Transition Used}
\label{sec: transitions}

\begin{figure}
\center
\begin{tabular}{@{}ccc@{}}
\includegraphics[scale=0.3, angle=0]{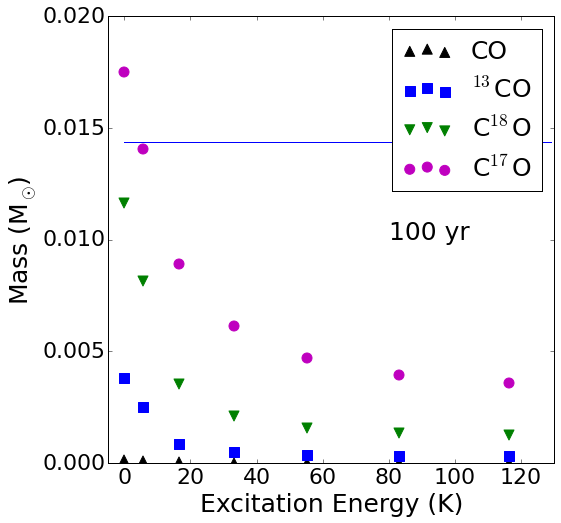} &
\includegraphics[scale=0.3, angle=0]{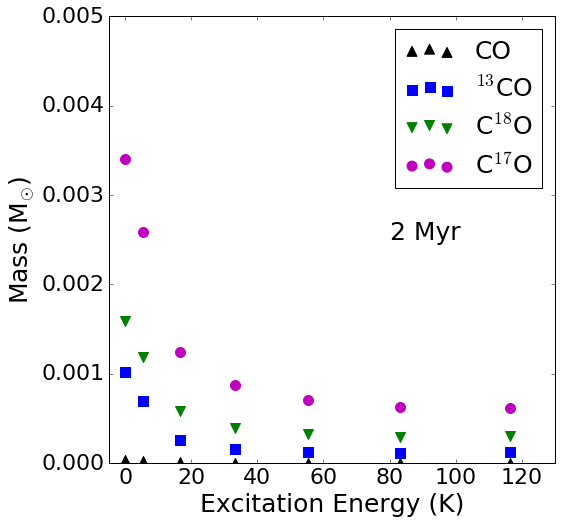} &
\includegraphics[scale=0.3, angle=0]{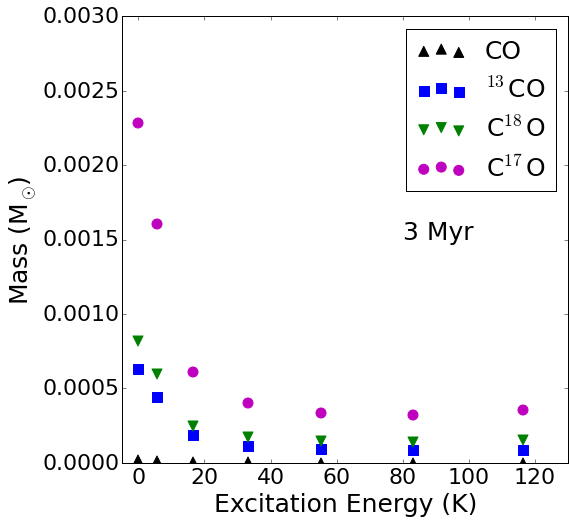} \\
\end{tabular}
 
\caption{The mass estimated from various transitions and different isotopologues. 
Excitation energies are for the upper excitation state of each transition. The 
populations at the zero excitation energy are extrapolated from the higher energy
populations with a three degree polynomial function. The actual disk mass is marked
by the blue line in the 100 yr diagram (on the left), and is above the chart in plots 
for the 2 Myr and 3 Myr disks because the mass is hugely underestimated with
the shown method. The actual disk masses in the input models at those three epochs
are $0.0144 \Msun$,  $0.0114 \Msun$ and $0.0107 \Msun$.
}
\label{mass_excitation}
\end{figure}

Because the temperature and optical depth are different in various parts 
of the disk, the mass estimation also depends on which isotopologue and 
which emission line is used. 
We show the mass estimated by different transitions and isotopologues at 
$100$ yr, $2$ Myr, and $3$ Myr of the disk evolution in Figure \ref{mass_excitation}. All
models assume optically thin emission for simplicity. The estimated 
mass decreases as we use higher-J lines with higher excitation energy for the mass
estimation. This is partially contributed by the large optical depth in the inner hot 
regions of the disk, partially because we are missing the low temperature CO that 
does not emit much at higher-J. 
Our models suggest that it is best to use lower $J$ transitions to 
measure disk mass.

\end{document}